\documentclass[prd,twocolumn,superscriptaddress,altaffilletter,nofootinbib]{revtex4}

\usepackage[dvips]{graphicx}
\usepackage{amsmath}
\usepackage{amssymb}
\usepackage{cancel}
\usepackage{graphicx,epsfig}
\newtheorem{theorem}{Theorem}
\newtheorem{corollary}{Corollary}[theorem]

\newcommand{\be}{\begin{equation}}
\newcommand{\ee}{\end{equation}}
\newcommand{\bea}{\begin{eqnarray}}
\newcommand{\eea}{\end{eqnarray}}
\newcommand{\der}{\partial}
\newcommand{\vphi}{\varphi}
\newcommand{\bet}{\begin{theorem}}
\newcommand{\eet}{\end{theorem}}
\newcommand{\bec}{\begin{corollary}}
\newcommand{\eec}{\end{corollary}}

\begin{document}



\title{Nonmetricity theories and aspects of gauge symmetry}



\author{Israel Quiros}\email{iquiros@fisica.ugto.mx}\affiliation{Dpto. Ingenier\'ia Civil, Divisi\'on de Ingenier\'ia, Universidad de Guanajuato, Gto., M\'exico.}



\begin{abstract}
In this paper we discuss on the phenomenological viability of nonmetricity theories of gravity which are based in the class of generalized Weyl spacetimes -- denoted by $W_4$ -- where arbitrary nonmetricity is allowed. This class of geometry includes the so called teleparallel spaces $Z_4$, which are the geometric basement of the symmetric teleparallel theories (STTs). The guiding principle in our discussion is Weyl gauge symmetry (WGS), which is a manifest symmetry of $W_4$ spaces. Here we derive the master equation that drives the gauge invariant variations of the length of vectors during parallel transport in $W_4$. This is the mathematical basis of the second clock effect (SCE). We are able to give qualitative and quantitative estimates for the SCE, as well as for the perihelion shift, in the coincident gauge of $Z_4$ space. We conclude that generalized Weyl spaces do not represent phenomenologically viable descriptions of Nature due to the SCE and, also to their predictions for the perihelion shift. All of the present results are based in the assumption of: (i) a gauge invariant parallel transport law and (ii) a consistency hypothesis which enables identifying hypothetical vectors and tensors defined in $W_4$, with related physical vectors and tensors arising in the given gravitational theory. Our discussion is mostly geometrical without relying on specific theories of gravity, unless it is absolutely necessary. 
\end{abstract}



\maketitle


\section{Introduction}\label{intro} 


In the search for new theoretical frameworks where to settle the present mysteries of our Universe, a resurgence of old ideas about non Riemannian geometry \cite{weyl-1917, dirac-1973, utiyama-1973, maeder-1978, cartan-1923, cartan-1924, einstein-1928, hehl-1976, hayashi-1979} has been evident in recent dates. In particular torsion theories based either in Riemann-Cartan space $U_4$ \cite{hehl-1976} or in its subclass known as Weitzenb$\ddot{\text o}$ck space $A_4$ \cite{hayashi-1979} whose connection has vanishing curvature, have been intensively studied. Theories that are based in Weitzenb$\ddot{\text o}$ck space $A_4$ are known as teleparallel theories of gravity \cite{hammond-2002, pereira-2004, pereira-1997, pereira-arxiv, pereira-2003, pereira-book, barrow-2011, sari-2011, leon-2012, maluf-prd-2012, maluf-andp-2012, maluf-2013, momeni-2014, silva-2016, silva-2017, capozziello-2016, golovnev-2017, coley-2019, pereira-2015, beltran-univ-2019, hohmann-2021, bahamonde-2021}. In this new wave of post-Riemannian theories a branch known as nonmetricity theories has recently arisen. To this branch belong the so called symmetric teleparallel theories \cite{delhom-2019, nester, adak-2006, adak-2006-1, adak-2013, beltran-plb-2016, javr-prd-2018, vilson-prd-2018, formiga-2019}, which have found interesting applications in the cosmological context \cite{lavinia-review, beltran-prd-2020, saridakis-prd-2020, lazkoz-prd-2019, lazkoz-prd-2021, sanjay-prd-2020, sanjay-2021}. 

Nonmetricity theories are formally presented as theories that are based in spacetimes whose affine properties differ from those of Riemann geometry, in particular due to nonvanishing nonmetricity. In other words, these are gravitational theories that are built over generalized Weyl geometry spaces. Actually, while standard Weyl spaces are characterized by the vectorial nonmetricity \cite{delhom-2019}, 

\bea \nabla_\alpha g_{\mu\nu}=-Q_\alpha g_{\mu\nu},\label{vect-nm}\eea where $Q_\alpha$ is the Weyl gauge vector and the covariant derivative $\nabla_\alpha$ is defined with respect to the affine connection of the manifold $\Gamma^\alpha_{\;\;\mu\nu}$, in generalized Weyl spaces the product of the gauge vector $Q_\alpha$ by the metric is replaced by generalized (arbitrary) nonmetricity:

\bea \nabla_\alpha g_{\mu\nu}=-Q_{\alpha\mu\nu}.\label{n-metricity}\eea The nonmetricity tensor $Q_{\alpha\mu\nu}$ is symmetric in the second and third indices. It measures how much the length of given vector varies during parallel transport \cite{beltran-univ-2019}. In consequence one may wonder whether these theories are affected as well by the second clock effect \cite{adler-book, perlick-1991, novello-1992, tucker, 2clock-1, 2clock-2, 2clock-3, 2clock-4, 2clock-5, 2clock-delhom, 2clock-no-1, 2clock-hobson, 2clock-tomi, tomi-replay}. The SCE was the reason why the original Weyl ideas about gauge invariant theory of gravity based in Weyl geometry spaces \cite{weyl-1917} were rejected in the first place \cite{scholz-2017}. Despite that there are scattered works that challenge the SCE \cite{2clock-no-1, 2clock-hobson, 2clock-tomi, tomi-replay}, the overwhelming majority of authors converge in the inevitability of this effect in Weyl geometry frameworks. There are a few works where the SCE is investigated within the framework of nonmetricity theories as well as of the STTs \cite{2clock-delhom, 2clock-tomi}, so that we feel this issue requires of more detailed investigation. 

A related aspect of generalized Weyl geometry that is avoided in papers on nonmetricity theories, including STTs and their cosmological applications, is the one about Weyl gauge symmetry which is tightly linked with the occurrence of the SCE \cite{tucker}. Given that WGS is a manifest symmetry of standard Weyl geometry spaces, which are distinguished by vectorial nonmetricity \eqref{vect-nm}, one may wonder; would WGS be a symmetry of generalized Weyl spaces as well? Would then the SCE arise also in spaces with arbitrary nonmetricity?

The above questions will be addressed in this paper, although these have been investigated before. For instance, in \cite{2clock-tomi} it has been recognized that the second clock effect might be an issue in the teleparallel geometric setup. In that reference, as a consequence of their lemma 2, the authors conclude that parallel transported objects in teleparallel spacetimes (symmetric or otherwise) do not experience a SCE (see also \cite{beltran-univ-2019}). It has been shown, also, that the geometrical foundation of 'purified gravity' \cite{beltran-prd-2018, bj-2019} is a generalisation of a Weyl integrable geometry (WIG). Hence, we found interesting to investigate these related issues from a gauge symmetric perspective: The guiding principle of our reasoning line will be Weyl gauge symmetry, which is the most salient property of Weyl spaces. We shall confirm a previous result \cite{delhom-2019} showing that WGS is also a manifest symmetry of generalized Weyl spaces. Our goal, then, is to search for the consequences of WGS within the framework of generalized Weyl geometry spaces, including the teleparallel spaces which are the geometric basement of the STTs \cite{beltran-univ-2019}. As we shall show, due to the SCE and to predictions for the perihelion shift that differ from general relativity (GR) results, arbitrary nonmetricity does not provide phenomenologically viable geometrical description of Nature. These results are based in the assumption of a specific gauge invariant parallel transport postulate (see equation \eqref{ptp} below) and of a consistency hypothesis that allows to identify physical vectors and tensors with the related (hypothetical) vectors and tensors defined in generalized Weyl space $W_4$.  

We have organized this paper in the following way. In the first ``mathematical'' part (sections \ref{sect-fundament}-\ref{sect-geod}) we expose the fundamentals of the mathematical machinery required to describe and understand the geometrical and physical (phenomenological) implications of arbitrary nonmetricity, that is exposed in the second ``physical'' part (sections \ref{sect-2clock}-\ref{sect-fermions}). In sections \ref{sect-fundament} and \ref{sect-telep} the basic notions of generalized Weyl geometry and the definition of teleparallel Weyl spacetime, respectively, are given. One of the most important subjects of our discussion: Weyl gauge symmetry, is exposed in section \ref{sect-conf-t}, while in subsection \ref{subsect-ginv-z4}, we explain why this symmetry is important for the symmetric teleparallel theories of gravity as well. In sections \ref{sect-gaug-der}, \ref{sect-p-transp}, \ref{sect-ginv-length} and \ref{sect-geod}, we introduce the notion of gauge derivative, gauge invariant parallel transport and geodesic and autoparallel equations, respectively. These subjects are indispensable to demonstrate the inevitability of the second clock effect in generalized Weyl spaces $W_4$ and, also, to be able to give qualitative and quantitative estimates of the perihelion shift. In particular, in section \ref{sect-ginv-length}, the master equations that drive the SCE are derived. The SCE is discussed in section \ref{sect-2clock} where, in subsection \ref{subsect-sce-z4} we give qualitative and quantitative estimates of this effect in the coincident gauge of teleparallel space $Z_4$. In the same framework we compute the amount of perihelion shift in $Z_4$. Qualitative and quantitative estimates of the perihelion shift are given in section \ref{sect-tests}. Then, in section \ref{sect-sce-challenge} we discuss on recent attempts at challenging the occurrence of the SCE in Weyl geometry spaces, while in section \ref{sect-postu} the relevance of the postulates that underlay the generalized Weyl geometrical setup is exposed. The crucial question about the possibility that fermions and other fields with the mass interact with the nonmetricity is handled in section \ref{sect-fermions}. Discussion of our results and conclusions are given in section \ref{sect-discuss}. At the end of the paper, for completeness, we have included an appendix section \ref{app-a} where the derivation of the equations of motion of particles with mass moving in $W_4$, is presented. Throughout the paper we use the metric with signature $(-+++)$ and, unless otherwise stated, the units system where $c=\hbar=1$.


\section{Fundamentals of generalized Weyl geometry}\label{sect-fundament}


We define the generalized Weyl spacetime, $W_4$, as the class of four-dimensional manifolds ${\cal M}_4$ that are paracompact, Hausdorff, connected $C^\infty$, endowed with a locally Lorentzian metric $g$ and the generalized linear (torsionless) affine connection $\Gamma$, that admits the following decomposition \cite{capozziello-2016, beltran-prd-2018, javr-prd-2018, vilson-prd-2018, beltran-prd-2020, poplawski}:

\bea \Gamma^\alpha_{\;\;\mu\nu}=\{^\alpha_{\mu\nu}\}+L^\alpha_{\;\;\mu\nu}\;\stackrel{\text{symb.}}{\longrightarrow}\;\Gamma=\{\}+L,\label{gen-aff-c}\eea where 

\bea \{^\alpha_{\mu\nu}\}:=\frac{1}{2}g^{\alpha\lambda}\left(\der_\nu g_{\mu\lambda}+\der_\mu g_{\nu\lambda}-\der_\lambda g_{\mu\nu}\right),\label{lc-aff-c}\eea is the Levi-Civita (LC) connection, also known as the Christoffel symbols of the metric, and

\bea L^\alpha_{\;\;\mu\nu}:=\frac{1}{2}\left(Q^{\;\;\alpha}_{\mu\;\;\nu}+Q^{\;\;\alpha}_{\nu\;\;\mu}-Q^\alpha_{\;\;\mu\nu}\right),\label{disformation}\eea is the disformation tensor. This is defined through the nonmetricity tensor $Q_{\alpha\mu\nu}$ in \eqref{n-metricity}, where $\nabla_\alpha$, denotes covariant derivative defined with respect to the generalized connection $\Gamma$ given by \eqref{gen-aff-c}. Alternatively, a hat over an object/operator means that it is defined in terms of the LC connection \eqref{lc-aff-c} instead. 

In this paper we call as ``generalized curvature tensor'' of $W_4$ spacetime the curvature of the connection ${\bf R}(\Gamma)$, whose coordinte components are defined in the following way:

\bea &&R^\alpha_{\;\;\sigma\mu\nu}:=\der_\mu\Gamma^\alpha_{\;\;\nu\sigma}-\der_\nu\Gamma^\alpha_{\;\;\mu\sigma}\nonumber\\
&&\;\;\;\;\;\;\;\;\;\;\;\;\;\;\;+\Gamma^\alpha_{\;\;\mu\lambda}\Gamma^\lambda_{\;\;\nu\sigma}-\Gamma^\alpha_{\;\;\nu\lambda}\Gamma^\lambda_{\;\;\mu\sigma},\label{gen-curv-t}\eea or, if take into account the decomposition \eqref{gen-aff-c}:

\bea &&R^\alpha_{\;\;\sigma\mu\nu}=\hat R^\alpha_{\;\;\sigma\mu\nu}+\hat\nabla_\mu L^\alpha_{\;\;\nu\sigma}-\hat\nabla_\nu L^\alpha_{\;\;\mu\sigma}\nonumber\\
&&\;\;\;\;\;\;\;\;\;\;\;\;\;\;+L^\alpha_{\;\;\mu\lambda}L^\lambda_{\;\;\nu\sigma}-L^\alpha_{\;\;\nu\lambda}L^\lambda_{\;\;\mu\sigma},\label{gen-curv-t-1}\eea where $\hat R^\alpha_{\;\;\sigma\mu\nu}$ is the Riemann-Christoffel or LC curvature tensor,

\bea &&\hat R^\alpha_{\;\;\sigma\mu\nu}:=\der_\mu\{^\alpha_{\nu\sigma}\}-\der_\nu\{^\alpha_{\mu\sigma}\}\nonumber\\
&&\;\;\;\;\;\;\;\;\;\;\;\;\;\;\;+\{^\alpha_{\mu\lambda}\}\{^\lambda_{\nu\sigma}\}-\{^\alpha_{\nu\lambda}\}\{^\lambda_{\mu\sigma}\},\label{lc-curv-t}\eea and $\hat\nabla_\alpha$ is the LC covariant derivative. Besides, the LC Ricci tensor $\hat R_{\mu\nu}=\hat R^\lambda_{\;\;\mu\lambda\nu}$ and LC curvature scalar read:

\bea &&\hat R_{\mu\nu}=\der_\lambda\{^\lambda_{\nu\mu}\}-\der_\nu\{^\lambda_{\lambda\mu}\}+\{^\lambda_{\lambda\kappa}\}\{^\kappa_{\nu\mu}\}-\{^\lambda_{\nu\kappa}\}\{^\kappa_{\lambda\mu}\},\nonumber\\
&&\hat R=g^{\mu\nu}\hat R_{\mu\nu},\label{lc-curv-sc}\eea respectively. We call $R^\alpha_{\;\;\sigma\mu\nu}$ as generalized curvature tensor because it is contributed both by LC curvature $\hat R^\alpha_{\;\;\sigma\mu\nu}$, and by nonmetricity through disformation $L^\alpha_{\;\;\mu\nu}$. We have that,

\bea &&R_{\mu\nu}=\hat R_{\mu\nu}+\hat\nabla_\lambda L^\lambda_{\;\;\mu\nu}-\hat\nabla_\nu L^\lambda_{\;\;\lambda\mu}\nonumber\\
&&\;\;\;\;\;\;\;\;\;\;\;+L^\lambda_{\;\;\lambda\kappa}L^\kappa_{\;\;\mu\nu}-L^\lambda_{\;\;\nu\kappa}L^\kappa_{\;\;\lambda\mu},\label{gen-ricci-t}\\
&&R=\hat R+{\cal Q}+\der{\cal Q},\label{gen-curv-sc}\eea where the nonmetricity scalar ${\cal Q}$ and the boundary term $\der{\cal Q}$, are defined as it follows:

\bea &&{\cal Q}:=L^\tau_{\;\;\tau\lambda}L^{\lambda\kappa}_{\;\;\;\;\kappa}-L^{\tau\kappa\lambda}L_{\lambda\tau\kappa},\nonumber\\
&&\der{\cal Q}:=\hat\nabla_\lambda\left(L^{\lambda\kappa}_{\;\;\;\;\kappa}-L^{\;\;\kappa\lambda}_\kappa\right).\label{q-inv}\eea 

Standard Weyl geometry space which is characterized by vectorial nonmetricity \eqref{vect-nm} and its particular case, known as Weyl integrable geometry (WIG), which is characterized by the choice $Q_\alpha=\der_\alpha\phi$ ($\phi$ is the Weyl gauge scalar), are subclasses in $W_4$. Here we shall denote these subclasses by $\tilde W_4$ and $i\tilde W_4$, respectively.


\subsection{Several properties and identities of the curvature in $W_4$}


For any torsionless connection $\nabla$ of $W_4$ space it is verified the (second) Bianchi identity:

\bea \nabla_\mu R^\kappa_{\;\;\lambda\nu\sigma}+\nabla_\nu R^\kappa_{\;\;\lambda\sigma\mu}+\nabla_\sigma R^\kappa_{\;\;\lambda\mu\nu}=0.\label{2-bianchi}\eea From this identity, taking into account that $$\nabla_\alpha g_{\mu\nu}=-Q_{\alpha\mu\nu},\;\nabla_\alpha g^{\mu\nu}=Q_\alpha^{\;\;\mu\nu},\;Q^\mu_{\;\;\mu\nu}=Q_\nu,$$ etc., and following a standard procedure, we obtain the following equation:

\bea &&\nabla^\nu G_{\nu\alpha}=\frac{1}{2}\left(Q_\alpha g^{\mu\nu}-Q_\alpha^{\;\;\mu\nu}\right)R_{\mu\nu}\nonumber\\
&&\;\;\;\;\;\;\;\;\;\;\;\;\;\;\;+\frac{1}{2}\left(Q_\lambda^{\;\;\mu\nu}-Q_\lambda g^{\mu\nu}\right)R^\lambda_{\;\;\mu\alpha\nu},\label{2-bianchi-einstein-t}\eea where $G_{\mu\nu}\equiv R_{\mu\nu}-g_{\mu\nu}R/2$ is the generalized Einstein's tensor. This equation amounts to a generalization of the Bianchi identity of the Einstein's tensor. In spacetimes endowed with standard Weyl geometry, since $$Q_\alpha g^{\mu\nu}-Q_\alpha^{\;\;\mu\nu}=\left(Q_\alpha-Q_\alpha\right)g^{\mu\nu}=0,$$ the well-known Bianchi identity of the generalized Einstein's tensor: $\nabla^\nu G_{\nu\alpha}=0$, is recovered.

The symmetries of the generalized curvature tensor in generalized Weyl space $W_4$ differ from those in Riemann space $V_4$. It is verified that:

\bea &&R^\alpha_{\;\;\sigma\mu\nu}=-R^\alpha_{\;\;\sigma\nu\mu},\label{symm-1-gcurv}\\
&&R_{\alpha\sigma\mu\nu}=-R_{\sigma\alpha\mu\nu}+\nabla_\mu Q_{\nu\alpha\sigma}-\nabla_\nu Q_{\mu\alpha\sigma}.\label{symm-2-gcurv}\eea The last equation is known as the third Biancchi identity and in compact form can be written in the following way \cite{2clock-tomi}:

\bea \nabla_{[\mu}Q_{\nu]\alpha\sigma}=R_{(\alpha\sigma)\mu\nu}.\label{3-bianchi}\eea Notice that, in standard Weyl geometry where $Q_{\alpha\mu\nu}=Q_\alpha g_{\mu\nu}$, so that $Q^\mu_{\;\;\mu\nu}=Q_\nu$ and $Q^{\;\;\mu}_{\nu\;\;\mu}=4Q_\nu$), 

\bea &&\nabla_\mu Q_{\nu\alpha\sigma}-\nabla_\nu Q_{\mu\alpha\sigma}=\left(\nabla_\mu Q_\nu-\nabla_\nu Q_\mu\right)g_{\alpha\sigma}\nonumber\\
&&\;\;\;\;\;\;\;\;\;\;\;\;\;\;\;\;\;\;\;\;\;\;\;\;\;\;\;\;\;\;\;\;=\left(\der_\mu Q_\nu-\der_\nu Q_\mu\right)g_{\alpha\sigma},\nonumber\eea the property \eqref{symm-2-gcurv} can be written in the following way:

\bea \der_{[\mu}Q_{\nu]}g_{\alpha\sigma}=R_{(\alpha\sigma)\mu\nu}.\label{3-bianchi-wig}\eea Besides, in the particular case when $Q_\alpha=\der_\alpha\phi$ ($\phi$ is a scalar function) -- known in the bibliography as WIG -- since $\der_\mu Q_\nu-\der_\nu Q_\mu=0$, the generalized curvature tensor possesses the same symmetries of the tensor indices as the Riemann-Christoffel curvature tensor.


\section{Teleparallel Weyl spacetime $Z_4$}\label{sect-telep}


A fully covariant metric-affine approach to the teleparallel formalism is based either on the following constraint on the generalized connection \cite{beltran-univ-2019, hohmann-2021}: 

\bea \Gamma^\alpha_{\;\;\mu\nu}=(\Lambda^{-1})^\alpha_{\;\;\lambda}\der_\mu\Lambda^\lambda_{\;\;\nu},\label{hohmann-c}\eea or, equivalently, on the teleparallel (flatness) requirement:

\bea {\bf R}(\Gamma)=0\;\Leftrightarrow\;R^\alpha_{\;\;\mu\sigma\nu}=0.\label{flat-r}\eea In equation \eqref{hohmann-c}, $\Lambda^\mu_{\;\;\nu}$ is an element of the general linear group $GL(4,\mathbb{R})$ \cite{beltran-univ-2019} and $(\Lambda^{-1})^\alpha_{\;\;\lambda}$ is its inverse, so that: $(\Lambda^{-1})^\mu_{\;\;\lambda}\Lambda^\lambda_{\;\;\nu}=\delta^\mu_\nu.$ The connection \eqref{hohmann-c} is purely inertial. In the absence of torsion this form of the connection leads to the additional constraint $\der_{[\mu}\Lambda^\alpha_{\;\;\nu]}=0$. Hence, the general element of $GL(4,\mathbb{R})$ determining the connection can be parametrized by a set of functions $\chi^\mu$ so that \cite{beltran-univ-2019}:

\bea \Gamma^\alpha_{\;\;\mu\nu}=\frac{\der x^\alpha}{\der\chi^\lambda}\der_\mu\der_\nu\chi^\lambda.\label{univ-2019}\eea

In what follows we shall call as teleparallel Weyl (geometry) space $Z_4$, a paracompact, Hausdorff, connected $C^\infty$ four-dimensional manifold ${\cal M}_4$, endowed with a locally Lorentzian metric $g$ and the generalized (torsionless) connection $\Gamma$ satisfying \eqref{univ-2019} and decomposed as in \eqref{gen-aff-c}:

\bea \Gamma^\alpha_{\;\;\mu\nu}=\{^\alpha_{\mu\nu}\}+L^\alpha_{\;\;\mu\nu}=\frac{\der x^\alpha}{\der\chi^\lambda}\der_\mu\der_\nu\chi^\lambda.\label{gww-connect}\eea 

Since, according to \eqref{flat-r}, the generalized curvature tensor of $Z_4$ vanishes:

\bea &&R^\alpha_{\;\;\sigma\mu\nu}=\der_\mu\Gamma^\alpha_{\;\;\nu\sigma}-\der_\nu\Gamma^\alpha_{\;\;\mu\sigma}\nonumber\\
&&\;\;\;\;\;\;\;\;\;\;\;\;\;+\Gamma^\alpha_{\;\;\mu\lambda}\Gamma^\lambda_{\;\;\nu\sigma}-\Gamma^\alpha_{\;\;\nu\lambda}\Gamma^\lambda_{\;\;\mu\sigma}=0,\label{vanish-curv-z4}\eea hence, the generalized Ricci tensor and the corresponding curvature scalar, both vanish as well:

\bea &&R_{\mu\nu}=\hat R_{\mu\nu}+\hat\nabla_\lambda L^\lambda_{\;\;\mu\nu}-\hat\nabla_\nu L^\lambda_{\;\;\lambda\mu}\nonumber\\
&&\;\;\;\;\;\;\;\;\;\;\;\;+L^\lambda_{\;\;\lambda\kappa}L^\kappa_{\;\;\mu\nu}-L^\lambda_{\;\;\nu\kappa}L^\kappa_{\;\;\lambda\mu}=0,\nonumber\\
&&R=\hat R+{\cal Q}+\der{\cal Q}=0,\label{weitz-rmn-r}\eea where we have taken into account equations \eqref{gen-ricci-t} and \eqref{gen-curv-sc}. The invariant scalar ${\cal Q}$ and the divergence term $\der{\cal Q}$, are given by \eqref{q-inv}. Equation \eqref{weitz-rmn-r} is the mathematical basis on which the claimed dynamical equivalence between GR and the symmetric teleparallel equivalent of general relativity (STEGR) holds \cite{beltran-univ-2019}.

The constraint \eqref{univ-2019} further restricts the fields $\chi^\mu$, the metric and the nonmetricity to satisfy (see \cite{beltran-univ-2019} for a similar relationship in the absence of nonmetricity): 

\bea \der_\alpha g_{\mu\nu}+Q_{\alpha\mu\nu}=2\frac{\der x^\lambda}{\der\chi^\sigma}\der_\alpha\der_{(\mu}\chi^\sigma g_{\nu)\lambda}.\label{telep-rel}\eea Under a specific choice of the field $\chi^\mu$, the above equation determines fixed relationships between the derivative of the metric and nonmetricity tensor. 

It is intuitive to notice that $Z_4\subset W_4$, so that all of the results obtained in $W_4$ will also apply to teleparallel Weyl space $Z_4$. As a matter of fact, $Z_4$ is equivalent to flat $W_4$ space:

\bea W_4\stackrel{{\bf R}=0}{\longrightarrow}Z_4.\label{w4-z4}\eea


\subsection{Coincident gauge of Teleparallel $Z_4$ space}\label{subsect-coinc-g}


Great simplification of computations is achieved if in \eqref{univ-2019} (also in \eqref{gww-connect}) set $\Gamma^\alpha_{\;\;\mu\nu}=0$. This choice is dubbed as ``coincident gauge'' since, vanishing of the connection can be interpreted as the gauge where the origin of the tangent space coincides with the spacetime origin \cite{beltran-prd-2018, beltran-univ-2019}. Vanishing of the affine connection in \eqref{gww-connect} leads to: 

\bea L^\alpha_{\;\;\mu\nu}=-\{^\alpha_{\mu\nu}\}\;\Rightarrow\;Q_{\alpha\mu\nu}=-\der_\alpha g_{\mu\nu}.\label{coinc-cond}\eea 

The latter ``coincident'' condition follows as well from the fact that we can completely remove the connection \eqref{univ-2019} by means of a diffeomorphism \cite{beltran-prd-2018}. Due to the right-hand equation in \eqref{coinc-cond}, the non-vanishing components of the nonmetricity tensor are known once the metric functions are given. Take, for instance, the static spherically symetric metric (we use spherical coordinates, $x^\mu=(t,r,\theta,\vphi)$):

\bea ds^2=-A^2dt^2+B^2dr^2+r^2d\theta^2+r^2\sin^2\theta d\vphi^2,\label{sp-symm-met}\eea where $A=A(r)$ and $B=B(r)$ are functions of the radial coordinate. In this case the non-vanishing components of non-metricity are:

\bea &&Q_{r00}=2AA_r,\;Q_{rrr}=-2BB_r,\;Q_{r\theta\theta}=-2r,\nonumber\\
&&\;Q_{r\vphi\vphi}=-2r\sin^2\theta,\;Q_{\theta\vphi\vphi}=-2r^2\sin\theta\cos\theta.\label{coinc-qabc}\eea Accordingly the only non-vanishing component of the vector $Q_\alpha\equiv Q^\mu_{\;\,\mu\alpha}$, is: $Q_r=-2B_r/B.$ 

Several of the computations in this paper will be performed in the coincident gauge for simplicity of handling and of the results.


\section{Weyl gauge symmetry}\label{sect-conf-t}


The Weyl rescalings or, also, Weyl gauge transformations, are composed of a conformal transformation of the metric:\footnote{As discussed in \cite{fulton_rmp_1962} conformal transformations can be formulated in different ways so that it is very important to distinguish these different formulations because they have different physical interpretation. In the present paper, in order to be specific, we assume that the conformal transformation of the metric within the Weyl rescalings does not represent a diffeomorphism or, properly, a conformal isometry. Moreover, the spacetime points -- same as spacetime coincidences or events -- as well as the spacetime coordinates that label the points in spacetime, are not modified or altered by the conformal transformations in any way.}

\bea g_{\mu\nu}\rightarrow\Omega^{-2}g_{\mu\nu},\label{scale-t}\eea where the positive smooth function $\Omega^2(x)$ is the conformal factor, plus simultaneous rescalings of the other fields $\Phi_i$ in the theory, according to their conformal weight $w$: $\Phi_i\rightarrow\Omega^w\Phi_i.$ 

The geometric laws that define $W_4$, among which is the nonmetricity condition \eqref{n-metricity}, are invariant under generalized Weyl gauge transformations. These amount to simultaneous conformal transformations of the metric \eqref{scale-t} and gauge transformations of nonmetricity \cite{delhom-2019, saridakis-prd-2020}:

\bea &&g_{\mu\nu}\rightarrow\Omega^2g_{\mu\nu},\;g^{\mu\nu}\rightarrow\Omega^{-2}g^{\mu\nu},\nonumber\\
&&Q^\alpha_{\;\;\mu\nu}\rightarrow Q^\alpha_{\;\;\mu\nu}-2\der^\alpha\ln\Omega\,g_{\mu\nu},\nonumber\\
&&Q_{\alpha\mu\nu}\rightarrow\Omega^2\left(Q_{\alpha\mu\nu}-2\der_\alpha\ln\Omega\,g_{\mu\nu}\right),\nonumber\\
&&Q_\alpha\rightarrow Q_\alpha-2\der_\alpha\ln\Omega,\label{gauge-t}\eea respectively, where we have defined 

\bea Q_\alpha:=Q^\lambda_{\;\;\lambda\alpha}=Q^\lambda_{\;\;\alpha\lambda}.\label{q-vec}\eea In what follows we shall call interchangeably the transformations \eqref{gauge-t} either as Weyl gauge transformations (WGT) or, simply, as gauge transformations. Although invariance and covariance under given transformations are quite different concepts, in this paper, often we shall use generically the word ``invariance'' to mean any of these different concepts.

Under \eqref{gauge-t} the LC connection and the disformation transform in the following way:

\bea &&\{^\alpha_{\mu\nu}\}\rightarrow\{^\alpha_{\mu\nu}\}+\left(\delta^\alpha_\mu\der_\nu+\delta^\alpha_\nu\der_\mu-g_{\mu\nu}\der^\alpha\right)\ln\Omega,\nonumber\\
&&L^\alpha_{\;\;\mu\nu}\rightarrow L^\alpha_{\;\;\mu\nu}-\left(\delta^\alpha_\mu\der_\nu+\delta^\alpha_\nu\der_\mu-g_{\mu\nu}\der^\alpha\right)\ln\Omega,\label{connect-t}\eea so that the generalized affine connection \eqref{gen-aff-c} is unchanged by the gauge transformations:

\bea \Gamma^\alpha_{\;\;\mu\nu}\rightarrow\Gamma^\alpha_{\;\;\mu\nu}.\label{aff-c-t}\eea This means that, under \eqref{gauge-t}: 

\bea R^\alpha_{\;\;\mu\sigma\nu}\rightarrow R^\alpha_{\;\;\mu\sigma\nu},\;R_{\mu\nu}\rightarrow R_{\mu\nu}\Rightarrow\;R\rightarrow\Omega^{-2}R.\label{curv-t-conf-t}\eea I. e., the generalized curvature tensor $R^\alpha_{\;\;\sigma\mu\nu}$ in \eqref{gen-curv-t} and the generalized Ricci tensor, $R_{\mu\nu}\equiv R^\lambda_{\;\;\mu\lambda\nu}$, are unchanged by \eqref{gauge-t}, while the generalized curvature scalar is transformed indeed.


\subsection{WGS and STTs}\label{subsect-ginv-z4}


Due to the claimed equivalence between GR and STEGR \cite{beltran-univ-2019}, the simplest of the symmetric teleparallel theories, it may be argued that WGS must not be a symmetry of STTs. In this regard let us point out that, on the one hand, the mentioned equivalence is based on equation \eqref{gen-curv-t-1} and the teleparallel requirement \eqref{flat-r}; $R^\alpha_{\;\;\sigma\mu\nu}=0$, which lead to the following equation:

\bea &&\hat R^\alpha_{\;\;\sigma\mu\nu}=\hat\nabla_\nu L^\alpha_{\;\;\mu\sigma}-\hat\nabla_\mu L^\alpha_{\;\;\nu\sigma}\nonumber\\
&&\;\;\;\;\;\;\;\;\;\;\;\;\;\;+L^\alpha_{\;\;\nu\lambda}L^\lambda_{\;\;\mu\sigma}-L^\alpha_{\;\;\mu\lambda}L^\lambda_{\;\;\nu\sigma}.\label{gen-curv-t-0}\eea From this equation it follows that (see equation \eqref{weitz-rmn-r}), 

\bea \hat R=-{\cal Q}-\der{\cal Q},\label{equiv}\eea where the nonmetricity scalar ${\cal Q}$ and the boundary term $\der{\cal Q}$ are defined in \eqref{q-inv}. On the other hand the teleparallel condition given by equations \eqref{hohmann-c}/\eqref{flat-r}, is invariant under \eqref{gauge-t}. Moreover, equations \eqref{gen-curv-t-1}, \eqref{gen-curv-t-0} and \eqref{equiv}, on which the mentioned equivalence between GR and STEGR is based, are gauge invariant as well. Take, for instance, the coincident gauge of teleparallel $Z_4$ space, where the connection vanishes \cite{beltran-univ-2019, adak-2013}: 

\bea \Gamma^\alpha_{\;\;\mu\nu}=0\;\Rightarrow\;\{^\alpha_{\mu\nu}\}=-L^\alpha_{\;\;\mu\nu},\label{coincident-cond}\eea i. e., we have that,

\bea \der_\alpha g_{\mu\nu}=-Q_{\alpha\mu\nu}.\label{coincident-cond-1}\eea Equations \eqref{coincident-cond} and \eqref{coincident-cond-1} are also invariant under \eqref{gauge-t}. 

We should not forget that the geometrical basis of STTs is flat $W_4$ space \eqref{w4-z4}, i. e., teleparallel $Z_4$ space, which is manifest gauge symmetric. Hence, one should expect that WGS, being a manifest symmetry of $Z_4$ spaces, should be shared at least as an implicit symmetry, by STTs and by GR. The latter theory may be understood as a particular gauge of a gauge invariant theory of gravity \cite{quiros-arxiv}. Hence, WGS should play a role in the development of the STTs as well. This has not been the case, since most of the STT Lagrangians of the form $f({\cal Q})$, that have been studied as models of our cosmos so far, do not respect gauge symmetry.\footnote{Consideration of given gravitational action $S_g=\int d^4x\sqrt{-g}{\cal L}_g$, where ${\cal L}_g$ is the gravitational Lagrangian density, adds additional possibilities. One may consider, for instance, a gauge invariant gravitational Lagrangian $\sqrt{-g}{\cal L}_g$, so that the derived gravitational equations will respect the manifest symmetry of background space $W_4$. Or one may, alternatively, consider a gravitational Lagrangian without gauge symmetry, even if the geometric background space $W_4$ is gauge symmetric. This, of course, will lead to a gravitational theory that is not gauge invariant. In this last case the manifest symmetry of the geometric background (gauge symmetry) is underutilized and may be ignored. In consequence, gauge invariant derivative operators may be replaced by non-gauge invariant ones: $\nabla^*_\alpha\rightarrow\nabla_\alpha$, $D^*/d\xi\rightarrow D/d\xi$, etc.} An exception may be the theories considered in \cite{saridakis-prd-2020}, where a family of conformal -- thus gauge invariant -- theories with second-order field equations and having the metric tensor as the fundamental variable, was formulated within the symmetric teleparallel framework. 

Although WGS should play a fundamental role in STTs, this assumption is clearly valid only in the class of theories explored in \cite{saridakis-prd-2020}.


\section{Gauge derivative}\label{sect-gaug-der}


In order to make the Weyl gauge symmetry compatible with well-known elementary derivation procedure and with the inclusion of fields into $W_4$, it is necessary to introduce the Weyl gauge derivative \cite{dirac-1973, utiyama-1973} (see also \cite{maeder-1978}.) 

Let ${\bf T}(w)$ be a tensor with coordinate components $T^{\alpha_1\alpha_2\cdots\alpha_i}_{\beta_1\beta_2\cdots\beta_j}$ and conformal weight $w$. I. e., under the gauge transformations \eqref{gauge-t}, it transforms as: ${\bf T}\rightarrow\Omega^w {\bf T}$. Then, the Weyl gauge differential of the object and, correspondingly, the Weyl gauge derivative, are defined as it follows (recall that $Q_\alpha:=Q^\lambda_{\;\;\lambda\alpha}$):

\bea &&d^*{\bf T}(w):=d{\bf T}(w)+\frac{w}{2}Q^*_\lambda dx^\lambda {\bf T}(w),\label{gauge-diff}\\
&&\der^*_\alpha{\bf T}(w):=\der_\alpha{\bf T}(w)+\frac{w}{2}Q^*_\alpha{\bf T}(w),\label{gauge-der}\eea where

\bea d^*{\bf T}=dx^\mu\der^*_\mu{\bf T},\label{rel-d-dm}\eea and

\bea Q^*_\alpha\equiv\frac{a}{s}Q_\alpha+\frac{b}{4s}Q_{\alpha\;\;\mu}^{\;\;\mu},\label{qaster}\eea is a linear combination of contributions $Q_\alpha\equiv Q^\mu_{\;\;\mu\alpha}$ and $Q_{\alpha\;\;\mu}^{\;\;\mu}$, with arbitrary constants $a$, $b$ and $s=a+b$. Notice that in standard Weyl space $\tilde W_4$, where $Q_{\alpha\mu\nu}=Q_\alpha g_{\mu\nu}$, since $Q_{\alpha\;\;\mu}^{\;\;\mu}=4Q_\alpha$, from \eqref{qaster} it follows that, 

\bea Q^*_\alpha=\frac{a+b}{s}Q_\alpha=Q_\alpha,\label{qaster-tilde-w4}\eea so that $Q^*_\alpha$ coincides with the Weyl gauge vector.

The above definitions warrant that both, the gauge differential and the gauge derivative, transform like the geometrical object itself, i. e., under the gauge transformations \eqref{gauge-t}:

\bea d^*{\bf T}(w)\rightarrow\Omega^wd^*{\bf T}(w),\;\der^*_\alpha{\bf T}(w)\rightarrow\Omega^w\der^*_\alpha{\bf T}(w).\nonumber\eea

Let us underline that the above definition of gauge derivative, is the basis for the definition of the gauge covariant derivative (called co-covariant derivative in \cite{dirac-1973}). Actually, if in \eqref{gauge-der} replace the partial derivative by covariant derivative operator $\der_\mu\rightarrow\nabla_\mu$, we obtain the expression for the gauge covariant derivative:

\bea \nabla^*_\alpha{\bf T}:=\nabla_\alpha{\bf T}+\frac{w}{2}Q^*_\alpha{\bf T}.\label{gauge-cov-der}\eea 

The use of gauge covariant derivative in standard Weyl geometry space $\tilde W_4$ allows to write the given equations in more compact form. For instance, the standard Weyl (vectorial) nonmetricity:

\bea \nabla_\alpha g_{\mu\nu}=-Q_\alpha g_{\mu\nu},\label{weyl-nm}\eea can be written as

\bea \nabla^*_\alpha g_{\mu\nu}=0.\label{co-cov}\eea However, both equations \eqref{weyl-nm} and \eqref{co-cov} are gauge covariant, so that the choice of the gauge covariant derivative instead of just the covariant derivative, in this case is only for purpose of compactness of writing, no more. Moreover, in generalized Weyl space $W_4$, where \eqref{n-metricity} takes place, the use of the gauge covariant derivative does not simplify writing in this case, since \eqref{n-metricity} is replaced by:

\bea \nabla^*_\alpha g_{\mu\nu}=Q^*_\alpha g_{\mu\nu}-Q_{\alpha\mu\nu},\label{n-met}\eea where in \eqref{gauge-der} we took into account that the conformal weight of the metric $w({\bf g})=2$. We may define, as well, the gauge covariant derivative of the tensor ${\bf T}$ along the wordline $x^\mu(\xi)$:

\bea \frac{D^*{\bf T}}{d\xi}=\frac{dx^\mu}{d\xi}\nabla^*_\mu{\bf T}.\label{direct-cov-der}\eea


\section{Parallel transport in $W_4$}\label{sect-p-transp}


When discussing about length variations that take place in the framework of Weyl geometry, in standard textbooks -- see, for instance, \cite{adler-book} -- one encounters arguments like this: Given that under an infinitesimal parallel transport the components $v^\alpha$ of a given vector ${\bf v}$ change according to

\bea dv^\alpha=-\Gamma^\alpha_{\;\;\mu\nu}v^\mu dx^\nu,\label{parallel-t}\eea then, in $W_4$ the length squared of the vector $||{\bf v}||^2=g_{\mu\nu}v^\mu v^\nu$, changes under parallel transport:

\bea d||{\bf v}||^2=\nabla_\lambda g_{\mu\nu}v^\mu v^\nu dx^\lambda=-Q_{\lambda\mu\nu}v^\mu v^\nu dx^\lambda.\label{length-parallel-t}\eea In the particular case of standard Weyl geometry \cite{weyl-1917, dirac-1973, utiyama-1973, maeder-1978, perlick-1991, scholz-2017, quiros-grg}, where vectorial nonmetricity \eqref{vect-nm} takes place: $Q_{\alpha\mu\nu}=Q_\alpha g_{\mu\nu}$, and assuming that the vector is transported from point $x_0$ to point $x$ along the curve $\cal C$, integration of this equation leads to: 

\bea v(x)=v_0\exp{\left(-\frac{1}{2}\int_{\cal C} Q_\alpha dx^\alpha\right)},\label{length-change}\eea where we adopted the notation $v(x)\equiv||{\bf v}(x)||$ and $v_0$ is an integration constant that is not transformed by the gauge transformations \eqref{gauge-t}. We identify this integration constant with the length of vector ${\bf v}$ at the starting point of the parallel transport trajectory: $v_0=||{\bf v}(x_0)||$. 

The line integral in \eqref{length-change} depends on followed path. The latter equation has a serious drawback: under the gauge transformations \eqref{gauge-t}, in particular since $Q_\alpha\rightarrow Q_\alpha-2\der_\alpha\ln\Omega$, according to \eqref{length-change}, the magnitude of the vector ${\bf v}$ transforms like: $v\rightarrow\Omega v,$ so that, under the assumption of gauge symmetry, the law \eqref{length-change} would be valid only for vectors with vanishing conformal weight $w=0$. It is not valid for arbitrary vectors ${\bf z}$ with $w({\bf z})=w$, since in general (arbitrary non-vanishing weight $w\neq 0$), under \eqref{gauge-t}: $$z^\alpha\rightarrow\Omega^w z^\alpha\;\Rightarrow\;z=\sqrt{g_{\mu\nu}z^\mu z^\nu}\rightarrow\Omega^{w+1} z.$$ For instance, tangent vectors ${\bf t}$: $t^\alpha=dx^\alpha/d\xi$ ($\xi$ is a parameter along the given curve), whose conformal weight $w({\bf t})=-1$, have weightless length $t\equiv||{\bf t}||$, which means that, under the gauge transformations \eqref{gauge-t}: $t\rightarrow t$. In particular, for the four-velocity ${\bf u}$ with components $u^\alpha$, whose length -- according to our metric signature choice -- is the imaginary unity $u=i$, one should have that:

\bea u(P)=u(0)=i,\label{unit-length}\eea at any point in $W_4$ spacetime. This would lead to $\int_{\cal C} Q_\alpha dx^\alpha=0$, which is true only if $Q_\alpha=0$, i. e., in Riemannian background spacetimes. There are other vectors, such as, for instance the four-momentum ${\bf p}$, with components $p^\alpha=mdx^\alpha/ids$ whose conformal weight $w({\bf p})=-2$, so that, under \eqref{gauge-t} its length squared transforms according to: $p^2\rightarrow\Omega^{-2} p^2$. Hence, equations \eqref{parallel-t}, \eqref{length-parallel-t} and \eqref{length-change} must be modified in such a way that they be gauge invariant and, consequently, weight-dependent. 

Below we shall follow a program similar to that of \cite{2clock-no-1}, where it is proposed to adopt the gauge covariant derivative identified in the geometric interpretation of Weyl gauge theories \cite{dirac-1973, utiyama-1973, maeder-1978} and to properly take into account the scaling dimension (conformal weight) of physical quantities.


\subsection{Gauge invariant parallel transport}\label{subsect-ginv-p-transp}


The above discussed drawback of equations like \eqref{parallel-t}, \eqref{length-parallel-t} and \eqref{length-change}, can be avoided if take into account the manifest gauge symmetry of $W_4$ space. Under this assumption the introduction of gauge differential \eqref{gauge-diff}, gauge derivative \eqref{gauge-der}, gauge covariant derivative, etc. is mandatory \cite{dirac-1973, utiyama-1973, maeder-1978}. 

Let ${\cal C}$ be a curve parametrized by the affine parameter $\xi$: $x^\mu(\xi)$. We can define the gauge covariant derivative along the path $x^\mu(\xi)$ to be expressed by the following operator:

\bea \frac{D^*}{d\xi}:=\frac{dx^\mu}{d\xi}\nabla^*_\mu,\label{gauge-cov-der-path}\eea where the gauge covariant derivative $\nabla^*_\mu$ is given by \eqref{gauge-cov-der}. Then, the parallel transport of given tensor ${\bf T}$ with coordinate components $T^{\alpha_1\alpha_2\cdots\alpha_i}_{\beta_1\beta_2\cdots\beta_j}$, along the path $x^\mu(\xi)$, is defined by the following requirement:

\bea \frac{D^*{\bf T}}{d\xi}:=\frac{dx^\mu}{d\xi}\nabla^*_\mu{\bf T}=0.\label{ptp}\eea The components of the tensor themselves are unchanged during parallel transport along the worldline ${\cal C}$:

\bea \frac{D^*}{d\xi}T^{\alpha_1\alpha_2\cdots\alpha_i}_{\beta_1\beta_2\cdots\beta_j}=\frac{dx^\mu}{d\xi}\nabla^*_\mu T^{\alpha_1\alpha_2\cdots\alpha_i}_{\beta_1\beta_2\cdots\beta_j}=0.\label{parallel-t-tensor-c}\eea This definition of gauge invariant parallel transport of a tensor in $W_4$, is not valid for tangent vectors of weight $w=-1$ as we shall show below. 

The fact that tangent unit vectors with weight $w=-1$ (both, timelike as the four-velocity $u^\mu$ and spacelike as $t^\mu=v^\mu/v$) do not obey \eqref{ptp}, allows derivation of equation of variation of length of given vectors under parallel transport. Consider, for instance, the tangent unit vector $\vec{\tau}$ with weight $w(\vec{\tau})=-1$ and coordinate components $\tau^\mu$, such that $(\vec{\tau},\vec{\tau})=g_{\mu\nu}\tau^\mu\tau^\nu=\pm 1$. Then take the gauge covariant derivative of both sides of the latter equation along the worldline $x^\mu(\xi)$; $\frac{D^*g_{\mu\nu}}{d\xi}\tau^\mu\tau^\nu+2g_{\mu\nu}\tau^\nu\frac{D^*\tau^\mu}{d\xi}=0$. We get that

\bea \frac{D^*\tau^\alpha}{d\xi}=\frac{1}{2}\left(Q^{\;\;\alpha}_{\mu\;\;\nu}-Q^*_\mu\delta^\alpha_\nu\right)\tau^\nu\frac{dx^\mu}{d\xi}.\label{tangent-v-p-transport}\eea It is seen from this equation that in standard Weyl space $\tilde W_4$, since $Q_{\alpha\mu\nu}=Q_\alpha g_{\mu\nu}$ and $Q^*_\alpha=Q_\alpha$, any vectors, no matter what their weight is, obey the parallel transport law \eqref{ptp}. 

Since any vector ${\bf v}$ with components $v^\alpha$ and weight $w({\bf v})=w\neq-1$, can be written as ${\bf v}=v\vec{\tau}$, and since ${\bf v}$ obeys the law of parallel transport \eqref{ptp}, then: 

\bea \frac{D^*v^\alpha}{d\xi}=\frac{D^*v}{d\xi}\tau^\alpha+v\frac{D^*\tau^\alpha}{d\xi}=0,\nonumber\eea from where we found that

\bea \frac{1}{v}\frac{D^*v}{d\xi}=\mp\frac{1}{2}\left(\pm Q^*_\lambda-Q_{\lambda\mu\nu}\tau^\mu\tau^\nu\right)\frac{dx^\lambda}{d\xi},\label{cov-der-uvect}\eea where the upper sign is for spacelike vector, while the lower sign corresponds to a timelike vector instead. From this equation we can straightforwardly derive equation \eqref{master-eq-vec} below.


\section{Gauge invariant length variations}\label{sect-ginv-length}


Let ${\bf v}$ and ${\bf w}$ be vectors with coordinate components $v^\mu$, $w^\mu$ and with conformal weigths $w({\bf v})=w_{\bf v}\neq-1$ and $w({\bf w})=w_{\bf w}\neq-1$, respectively. Let these vectors to be parallel transported along the worldline $x^\mu(\xi)$. The gauge covariant derivative of their inner product $({\bf v},{\bf w})=g_{\mu\nu}v^\mu w^\nu$, vanishes during parallel transport:

\bea \frac{D^*({\bf v},{\bf w})}{d\xi}=\frac{D^*g_{\mu\nu}}{d\xi}v^\mu w^\nu=0,\label{gauge-cov-der-dot}\eea where we took into account that, since both ${\bf v}$ and ${\bf w}$ are parallel transported along the path ${\cal C}$, then: $D^*v^\mu/d\xi=0$, $D^*w^\mu/d\xi=0$. Besides, since on the one hand $D^*\psi/d\xi=D\psi/d\xi$, where $\psi$ is any scalar quantity, the weight of the scalar product: $w({\bf v},{\bf w})=2+w_{\bf v}+w_{\bf w}$, and taking into account \eqref{gauge-diff}, one gets the following equation:

\bea \frac{D^*({\bf v},{\bf w})}{d\xi}=\frac{d({\bf v},{\bf w})}{d\xi}+\left(\frac{2+w_{\bf v}+w_{\bf w}}{2}\right)Q^*_\mu\frac{dx^\mu}{d\xi}({\bf v},{\bf w}).\nonumber\eea On the other hand,

\bea \frac{D^*g_{\mu\nu}}{d\xi}=\frac{dx^\alpha}{d\xi}\nabla^*_\alpha g_{\mu\nu}=\frac{dx^\alpha}{d\xi}\left(Q^*_\alpha g_{\mu\nu}-Q_{\alpha\mu\nu}\right).\nonumber\eea Substituting these equations into \eqref{gauge-cov-der-dot}, one obtains that:

\bea \frac{d\ln\psi}{d\xi}=-\left[\left(\frac{w_{\bf v}+w_{\bf w}}{2}\right)Q^*_\alpha+\frac{1}{\kappa}Q_{\alpha\mu\nu}t^\mu_{\bf v}t^\nu_{\bf w}\right]\frac{dx^\alpha}{d\xi},\label{master-eq}\eea where we have introduced the following notation: $\psi\equiv({\bf v},{\bf w})$, and 

\bea t^\mu_{\bf z}:=\frac{z^\mu}{z},\label{space-l-unit-v}\eea are the coordinate components of the spacelike unit vector ${\bf t}_{\bf z}:={\bf z}/z$ and $\kappa:=\cos\theta$ ($\theta$ is the angle formed among vectors ${\bf v}$ and ${\bf w}$). Besides, we took into account the following expression: $({\bf v},{\bf w})=\kappa vw.$ 

Notice that the quantity in square brackets in \eqref{master-eq} is not a partial derivative, so that $d\ln\psi$ is not a perfect differential. Equation \eqref{master-eq} can be formally integrated along the path ${\cal C}$ from the origin to the point $x$, to yield:

\bea \psi(x)=\psi(0)\;e^{-\int_{\cal C}\left[\left(\frac{w_{\bf v}+w_{\bf w}}{2}\right)Q^*_\alpha+\frac{1}{\kappa}Q_{\alpha\mu\nu}t^\mu_{\bf v}t^\nu_{\bf w}\right]dx^\alpha},\label{master-int}\eea where $\psi(0)$ is an integration constant that can be determined from the initial conditions. We recall that the above equations, as well as the equations below are valid only for vectors with weight $w\neq-1$.

If in \eqref{master-eq} replace the inner product of two vectors by the length squared of given vector, say of vector ${\bf v}$: $\psi=v^2=({\bf v},{\bf v})$, then

\bea \frac{d\ln v}{d\xi}=-\frac{1}{2}\left(w_{\bf v}Q^*_\alpha+Q_{\alpha\mu\nu}t^\mu_{\bf v}t^\nu_{\bf v}\right)\frac{dx^\alpha}{d\xi}.\label{master-eq-vec}\eea Formal integration of this equation leads to:

\bea v(x)=v(0)\;\exp{\left[-\frac{1}{2}\int_{\cal C}\left(w_{\bf v}Q^*_\alpha+Q_{\alpha\mu\nu}t^\mu_{\bf v}t^\nu_{\bf v}\right)dx^\alpha\right]}.\label{master-int-vec}\eea 

In standard Weyl geometry space $\tilde W_4$, with vectorial nonmetricity $Q_{\alpha\mu\nu}=Q_\alpha g_{\mu\nu}$, $Q_\alpha=Q^*_\alpha$, the above equation amounts to:

\bea v(x)=v(0)\;\exp{\left[-\frac{1+w_{\bf v}}{2}\int_{\cal C}Q_\mu dx^\mu\right]}.\label{master-int-vec-swg}\eea


\subsection{Path-dependent mass variation}


As we shall see, equation \eqref{master-int-vec} is the responsible for the SCE in generalized Weyl geometry space $W_4$. We have just to replace the arbitrary vector ${\bf v}$ by the four-momentum:

\bea {\bf p}=m{\bf u},\;p\equiv||{\bf p}||=\pm im,\label{4-p}\eea where due to our signature choice, $u\equiv||{\bf u}||=\pm i$, $m$ is the mass of the point-particle and ${\bf u}$ is the four-velocity, with coordinate components: $u^\mu=dx^\mu/d\tau$, $d\tau=ids$ is the proper time along the worldline ${\cal C}$ ($ds$ is the infinitesimal arc-length) and $i$ is the imaginary unit. The weight of the four-momentum is $w({\bf p})=-2$. 

If in \eqref{master-int-vec} make the replacement ${\bf v}\rightarrow{\bf p}$, one obtains the following equation driving the path-dependent variation of mass during parallel transport: 

\bea m(x)=m(0)\,\exp\int_{\cal C}\left(Q^*_\alpha+\frac{1}{2}Q_{\alpha\mu\nu}u^\mu u^\nu\right)dx^\alpha,\label{master-eq-mass}\eea where we took into account that: $$t^\mu_{\bf p}=\frac{p^\mu}{p}=\mp i\frac{dx^\mu}{d\tau}=\mp iu^\mu.$$ The mass parameter $m(x)$ is not properly a scalar field since it depends on followed path ${\cal C}$, hence it is uniquely defined at spacetime point $x$ once a path from $0$ to the point $x$ is specified. 

As seen from \eqref{master-eq-mass}, the magnitude of the mass, depends not only on followed path but also on speed $u^\mu$. The latter dependence is a new effect in generalized Weyl space $W_4$, with respect to $\tilde W_4$. Actually, in standard Weyl space $\tilde W_4$ the mass variation during parallel transport \eqref{master-eq-mass} reads:

\bea m(x)=m(0)\,\exp\left(\frac{1}{2}\int_{\cal C}Q_\mu dx^\mu\right).\label{master-eq-mass-swg}\eea This quantity only depends on path ${\cal C}$.

We want to conclude this section by noticing that equations like \eqref{master-int-vec} and \eqref{master-eq-mass} are direct consequence of the law of parallel transport \eqref{ptp} together with the nonmetricity law \eqref{n-metricity}. Hence, if one adopts generalized Weyl geometry space $W_4$ as the geometric setup of given gravitational theory, one adopts the occurrence of mass variation under parallel transport according to \eqref{master-eq-mass-swg}. One can not just say that, for instance, equation \eqref{master-eq-mass} is not valid, without geometrical consequences.


\section{Autoparallels and geodesics in Weyl space $W_4$}\label{sect-geod}


Within GR theory the gravitational effects are consequence of the LC curvature of spacetime ($\hat R^\alpha_{\;\;\sigma\mu\nu}\neq 0$) and the test particles move on time-like geodesics of $V_4$:

\bea \frac{d^2x^\alpha}{ds^2}+\{^\alpha_{\mu\nu}\}\frac{dx^\mu}{ds}\frac{dx^\nu}{ds}=0,\label{geod-eq}\eea where $ds^2=g_{\mu\nu}dx^\mu dx^\nu$ is the line element, which coincide with the ``straight lines'' of Riemann space. But, in general, autoparallels -- ``straightest curves'' of the geometry -- do not coincide with the geodesics, which are the ``shortest curves'' \cite{poplawski, adak-arxiv, obukhov}. 

There goes a discussion on whether autoparallels or geodesics describe the motion of test particles \cite{adak-arxiv, obukhov}. However, there are particular cases when autoparallels and geodesics coincide as, for instance, in GR. As we shall see, these coincide as well in $\tilde W_4$.

In general, the trajectories of spinor fields (like fermions) and of extended spinning test bodies in $W_4$ are neither autoparallels nor geodesics, which are valid only for spinless test (point) particles. While spinor fields satisfy the Dirac equation in curved background, pole-dipole particles and extended spinning test bodies obey the Mathisson-Papapetrou-Dixon equation \cite{mathisson, papapetrou, dixon, wald}.


\subsection{Geodesics}\label{subsect-geod}


In $V_4$ the equations \eqref{geod-eq} coincide with the equations of motion of point-like particles that can be derived from the action principle:

\bea S=m\int ds,\label{action-geod}\eea where the constant $m$ is the mass of test particles. In general those curves that solve the equations of motion are extremal curves, i. e., shortest or longest curves.  

In $W_4$ the length of vectors changes from point to pint, so that the mass parameter, being the length of the 4-momentum vector ($p^2=-m^2$), is not a constant anymore, but it is a function of the spacetime point: $m=m(x)$, as it can be seen from \eqref{master-eq-mass}. The modified action principle \eqref{action-geod} that operates in $W_4$, reads:

\bea S=\int mds,\label{action-geod-w4}\eea where, since under \eqref{scale-t}: $ds\rightarrow\Omega ds$ and $m\rightarrow\Omega^{-1}m$, the action is gauge invariant. The variational principle of least action applied to \eqref{action-geod-w4}, where $m=m(x)$ is a function of the spacetime point (also of followed path), leads to the following gauge covariant equations of motion that particles with the mass obey in $W_4$ (see appendix \ref{app-a} for a detailed derivation):

\bea \frac{d^2x^\alpha}{ds^2}+\{^\alpha_{\mu\nu}\}\frac{dx^\mu}{ds}\frac{dx^\nu}{ds}-\frac{\delta\ln m}{\delta x^\mu}h^{\mu\alpha}=0,\label{geod-w4}\eea where, 

\bea h^{\alpha\mu}=g^{\alpha\mu}-\frac{dx^\alpha}{ds}\frac{dx^\mu}{ds},\label{proj-t}\eea is the orthogonal projector tensor. It projects any vector or tensor onto the hypersurface which is orthogonal to the four-velocity ${\bf u}$. Equation \eqref{geod-w4} can be written in the following alternative form:

\bea &&\frac{d^2x^\alpha}{ds^2}+\Gamma^\alpha_{\;\;\mu\nu}\frac{dx^\mu}{ds}\frac{dx^\nu}{ds}-\frac{\delta\ln m}{\delta x^\mu}h^{\mu\alpha}\nonumber\\
&&\;\;\;\;\;\;\;\;\;\;-\left(Q^{\;\;\alpha}_{\mu\;\;\nu}-\frac{1}{2}Q^\alpha_{\;\;\mu\nu}\right)\frac{dx^\mu}{ds}\frac{dx^\nu}{ds}=0.\label{geod-w4-1}\eea

If one adopts $W_4$ space as the background where gravitational laws take place, then one should adopt \eqref{master-eq-mass} since, as shown in subsection \ref{subsect-ginv-p-transp}, \eqref{master-eq-mass} is consequence of our assumed definition of parallel transport law \eqref{ptp} and of equation \eqref{tangent-v-p-transport} which is, in turn, a direct consequence of the nonmetricity law \eqref{n-metricity} (if \eqref{n-metricity} is true, then \eqref{tangent-v-p-transport} is also true). In order to determine the last term in the left-hand side of \eqref{geod-w4} one needs an additional hypothesis or postulate.


\subsubsection{Consistency hypothesis}\label{subsect-consist-hypo}


Here we assume the following ``consistency'' hypothesis: The four-momentum of timelike test particles is to be identified with the hypothetical four-momentum \eqref{4-p} defined in $W_4$, i. e., the mass parameter $m$ appearing in \eqref{geod-w4} is to be identified with the mass in \eqref{master-eq-mass}. In consequence, the following equation:

\bea \frac{\delta\ln m}{\delta x^\alpha}=Q^*_\alpha-\frac{1}{2}Q_{\alpha\kappa\lambda}\frac{dx^\kappa}{ds}\frac{dx^\lambda}{ds},\label{mass-law'}\eea is to be substituted back into \eqref{geod-w4}. In restricted Weyl space $\tilde W_4$, since $Q_{\alpha\mu\nu}=Q_\alpha g_{\mu\nu}$, equation \eqref{mass-law'} can be written in a simplified form 

\bea \frac{\delta\ln m}{\delta x^\alpha}=\frac{1}{2}Q_\alpha.\label{mass-law}\eea The notation, $\delta/\delta x^\mu$, coming from the variational process -- see appendix \ref{app-a} --, has been used to underline that, in general, \eqref{mass-law'} is not a partial derivative since $d\ln m$ is not a perfect differential.

Equations \eqref{master-eq-mass}/\eqref{master-eq-mass-swg} and \eqref{mass-law'}/\eqref{mass-law} and the consistency hypothesis, are at the heart of the SCE \cite{adler-book, perlick-1991, novello-1992, tucker, 2clock-1, 2clock-2, 2clock-3, 2clock-4, 2clock-5, 2clock-delhom, 2clock-no-1, 2clock-hobson, 2clock-tomi, tomi-replay}. Non-vanishing of the ``variational gradient'' $\delta\ln m/\delta x^\alpha\neq 0$, is what makes the difference between the equation of motion of particles with mass in generalized Weyl space $W_4$ with respect to their motion in Riemann space $V_4$. Consequently, while Riemann geometry can not be associated with the second clock effect, under the assumption of gauge invariant parallel transport law \eqref{tangent-v-p-transport} and of the consistency hypothesis above stated, it is inevitable in generalized Weyl spaces. We may conclude that gauge invariance of \eqref{geod-w4} and the SCE are tightly linked.


\subsubsection{Null geodesics in $W_4$}\label{subsect-massless}


For photons and massless particles in general, the geodesic equations can be obtained from the following variational principle. Consider the action

\bea S_\text{ph}=\int{\cal L}(x^\mu,\dot x^\mu,\lambda)d\lambda,\label{ph-action}\eea where the dot means derivative with respect to the affine parameter $\lambda$ and ${\cal L}(x^\mu,\dot x^\mu,\lambda)=g_{\mu\nu}\dot x^\mu\dot x^\nu/2$. The path followed by photons is such that the variation: $\delta S_\text{ph}=0$. The resulting equations of motion coincide with the null geodesics of the Riemannian space:

\bea \frac{dk^\alpha}{d\lambda}+\{^\alpha_{\mu\nu}\}k^\mu k^\nu=0,\label{0-geod-w4}\eea where $k^\alpha\equiv dx^\alpha/d\lambda$ are the coordinate components of the wave vector of the photon ${\bf k}$.


\subsection{Autoparallels}


In generalized Weyl space $W_4$ the ``time-like'' autoparallels are those curves along which the tangent four-velocity vector ${\bf u}$ is parallel transported. However, as we have seen in subsection \ref{subsect-ginv-p-transp}, the unit tangent vectors with weight $w=-1$, like the four-velocity, do not satisfy the law of parallel transport \eqref{ptp}. In this case it is equation \eqref{tangent-v-p-transport} the one that takes place. Hence, for the four-velocity ${\bf u}$ with coordinate components $u^\mu$ and weight $w=-1$, we have that:

\bea \frac{D^*u^\alpha}{d\xi}=\frac{dx^\mu}{d\xi}\nabla^*_\mu u^\alpha=\frac{1}{2}\left(Q^{\;\;\alpha}_{\mu\;\;\nu}-Q^*_\mu\delta^\alpha_\nu\right)u^\nu\frac{dx^\mu}{d\xi}\neq 0.\nonumber\eea This equation can be written in more convenient forms:

\bea \frac{d^2x^\alpha}{ds^2}+\Gamma^\alpha_{\;\;\mu\nu}\frac{dx^\mu}{ds}\frac{dx^\nu}{ds}-\frac{1}{2}Q^{\;\;\alpha}_{\mu\;\;\nu}\frac{dx^\mu}{ds}\frac{dx^\nu}{ds}=0,\label{autop-w4}\eea or

\bea &&\frac{d^2x^\alpha}{ds^2}+\{^\alpha_{\mu\nu}\}\frac{dx^\mu}{ds}\frac{dx^\nu}{ds}\nonumber\\
&&\;\;\;\;\;\;\;\;\;\;+\frac{1}{2}\left(Q^{\;\;\alpha}_{\mu\;\;\nu}-Q^\alpha_{\;\;\mu\nu}\right)\frac{dx^\mu}{ds}\frac{dx^\nu}{ds}=0,\label{autop-w4-1}\eea where, without loss of generality, we have chosen the parameter along the autoparallel to be the arc-length $s$.


\subsubsection{Null autoparallels}


In the same fashion, in $W_4$ the ``null'' autoparallels are those curves along which the gauge covariant derivative of the wave vector ${\bf k}$ with components $k^\mu:=dx^\mu/d\lambda$ ($\lambda$ is a parameter along the null autoparallel), obeys:

\bea \frac{D^*k^\alpha}{d\lambda}=\frac{1}{2}\left(Q^{\;\;\alpha}_{\mu\;\;\nu}-Q^*_\mu\delta^\alpha_\nu\right)k^\mu k^\nu\neq 0,\nonumber\eea or

\bea &&\frac{dk^\alpha}{d\lambda}+\Gamma^\alpha_{\;\;\mu\nu}k^\mu k^\nu-\frac{1}{2}\left(Q^{\;\;\alpha}_{\mu\;\;\nu}+Q^*_\mu\delta^\alpha_\nu\right)k^\mu k^\nu=0,\nonumber\\
&&\frac{dk^\alpha}{d\lambda}+\{^\alpha_{\mu\nu}\}k^\mu k^\nu\nonumber\\
&&\;\;\;\;\;\;\;\;\;+\frac{1}{2}\left(Q^{\;\;\alpha}_{\mu\;\;\nu}-Q^\alpha_{\;\;\mu\nu}-Q^*_\mu\delta^\alpha_\nu\right)k^\mu k^\nu=0,\label{0-autop-w4}\eea where we have taken into account that the conformal weight of the wave vector $w({\bf k})=-2$, i. e., it coincides with the weight of the four-momentum since, in the quantum limit both should be related by ${\bf p}=\hbar{\bf k}$.\footnote{Here it is implicitly assumed that the Universal constant $\hbar$ is not transformed by the Weyl gauge transformations.}


\subsection{Geodesics and autoparallels in $\tilde W_4$}


In standard Weyl space since the vectorial nonmetricity \eqref{vect-nm} takes place and taking into account \eqref{mass-law}, it can be straightforwardly demonstrated that equations \eqref{geod-w4} and \eqref{autop-w4-1} coincide, so that, similar to what happens in Riemann space $V_4$, in $\tilde W_4$ time-like autoparallels and geodesic curves are one and the same thing. Actually, if set $Q_{\alpha\mu\nu}=Q_\alpha g_{\mu\nu}$ in the last term in \eqref{autop-w4-1}, one gets that:

\bea \frac{1}{2}\left(Q^{\;\;\alpha}_{\mu\;\;\nu}-Q^\alpha_{\;\;\mu\nu}\right)\frac{dx^\mu}{ds}\frac{dx^\nu}{ds}=-\frac{1}{2}Q_\mu h^{\mu\alpha},\nonumber\eea which coincides with the last term in \eqref{geod-w4} if take into account \eqref{mass-law}.

For the same reason, in standard Weyl geometry space $\tilde W_4$, for null-vector ${\bf k}$, since 

\bea g_{\mu\nu}k^\mu k^\nu=0\;\Rightarrow\;Q^\alpha_{\;\;\mu\nu}k^\mu k^\nu=0,\nonumber\eea the last term in the null-autoparallel equation \eqref{0-autop-w4} vanishes:

\bea \left(Q^{\;\;\alpha}_{\mu\;\;\nu}-Q^\alpha_{\;\;\mu\nu}-Q^*_\mu\delta^\alpha_\nu\right)k^\mu k^\nu=0,\nonumber\eea so that it coincides with the Riemannian null geodesic \eqref{0-geod-w4}. Hence, photons and massless particles probe the Riemannian (Levi-Civita) structure of spacetime. These do not interact with the vectorial nonmetricity. 

We want to underline that for arbitrary nonmetricity, since null-particles follow Riemannian null geodesics \eqref{0-geod-w4}, radiation and massless particles do not interact with nonmetricity.


\section{Second clock effect: a serious phenomenological issue}\label{sect-2clock}


We know that in general relativity, that is based on Riemann spacetimes, when two identical clocks, initially synchronized, are parallel transported along different paths, a certain loss of synchronization arises that is called as the ``first clock effect.'' In Weyl spacetimes an additional effect arises: the two clocks not only have lost their initial synchronization, but, they go at different rates. It is known as the ``second clock effect'' \cite{adler-book, perlick-1991, novello-1992, tucker, 2clock-1, 2clock-2, 2clock-3, 2clock-4, 2clock-5, 2clock-delhom, 2clock-no-1, 2clock-hobson, 2clock-tomi, tomi-replay}. In Weyl space $W_4$, including the subclasses $\tilde W_4$ and $Z_4$, the SCE causes a serious phenomenological and conceptual issue. In consequence, since WGS should be a implicit symmetry in the symmetric teleparallel framework, the issue must have impact, at least, in symmetric teleparallel theories of the kind explored in \cite{saridakis-prd-2020}, and also in most general gauge nonmetricity theories. This is not usually taken into account in the bibliography (as illustration see \cite{nester, adak-2006, adak-2006-1, adak-2013, beltran-plb-2016, javr-prd-2018, vilson-prd-2018, formiga-2019, beltran-prd-2020, saridakis-prd-2020, lazkoz-prd-2019, lazkoz-prd-2021, sanjay-prd-2020, sanjay-2021, beltran-univ-2019, helsinki, 2mann, frusciante, basilakos, lin, mandal, moraes, atayde, dambrosio, beltran-prd-2018, coincident-1, coincident-2} and related bibliographic references.) A similar issue in connection with the SCE was enough to dismiss the original Weyl's gauge invariant gravitational theory and its related geometrical framework \cite{weyl-1917}. 

In reference \cite{2clock-no-1} a different point of view is developed according to which the SCE does not take place. The outline of their argument is as follows. The weight of the four-velocity $u^\alpha(\lambda)$, where $\lambda$ is an arbitrary parameter along the worldline of the particle, is $w=-1$. The length of this vector is then invariant under Weyl gauge transformations. Moreover, by working in terms of the Weyl gauge covariant derivative, it is demonstrated that one may always find a parametrization for which the length of the tangent vector remains equal to unity under parallel transport along its worldline. Consequently, the original argument which leads to an SCE, is removed. A similar conclusion is reached in lemma 2 of reference \cite{2clock-tomi}. Critical comments on \cite{2clock-no-1} and \cite{2clock-tomi} are given in section \ref{sect-sce-challenge} and also in reference \cite{quiros-arxiv-2022}. 

Given that the parallel transport law \eqref{ptp} and the nonmetricity law \eqref{n-metricity} together, inevitable lead to equation \eqref{master-eq-mass}, then the lacking piece to relate the hypothetical SCE of geometric nature -- which is inherent in $W_4$ space -- with the physical SCE, is the ``reasonable'' consistency hypothesis \eqref{mass-law'}. It is of no relevance at all, whether autoparallels or geodesics are the physical trajectories of test particles. As we have shown in section \ref{sect-geod}, in none of these cases the four-velocity vector obeys the parallel transport law \eqref{ptp}. Besides, as already mentioned in several parts of the text, ``real'' test bodies either are extended objects or are spinning objects. For them it is the Mathisson-Papapetrou-Dixon equation the one that defines their dynamics in the gravitational field. Only scalar test point particles follow either autoparallels or geodesics. In subsection \ref{subsect-sce-z4} below, for simplicity of computations, we shall assume, precisely, the last case (scalar test particles). Besides, for definiteness, we assume that test particles follow geodesics \eqref{geod-w4} with the assumption that \eqref{mass-law'} is valid. In the concluding section \ref{sect-discuss} we shall come back to this subject.


\subsection{Atomic transitions and the SCE}


Below we shall show that, under the above reasonable assumptions, the SCE takes place in Weyl spaces $W_4$/$\tilde W_4$, as widely recognized. We shall base our discussion on the role of the mass parameter, given that one can carry an atomic clock which measures the International Atomic Time.\footnote{It is the primary international time standard from which other time standards are calculated.} The principle of operation of an atomic clock is based on atomic physics: it measures the electromagnetic signal that electrons in atoms emit when they change energy levels. For instance, the energy of each energy level in the hydrogen atom, labeled by $n$, is given by: 

\bea E_n\approx-\frac{m\alpha^2}{2n^2},\label{e-level}\eea where $m$ is the mass of the electron and $\alpha\approx 1/137$ is the fine-structure constant. Any changes in the mass $m$ over spacetime will cause changes in the energy levels and, consequently, in the energy of the atomic transitions 

\bea \omega_{if}=|E_{n_f}-E_{n_i}|=\frac{m\alpha^2}{2}\left(\frac{1}{n^2_f}-\frac{1}{n^2_i}\right).\label{atm-t}\eea Hence, the functioning of atomic clocks will be affected by the variation of masses over spacetime.

Contrary to such vectors like the four-velocity $u^\alpha=dx^\alpha/ids$, whose conformal weight is $w=-1$, hence its length $u=\pm i$, is gauge invariant, the mass is the length of the four-momentum ${\bf p}$ with components: $p^\alpha=mdx^\alpha/ids$. Its conformal weigth $w({\bf p})=-2$, so that $im=p=\sqrt{g_{\mu\nu}p^\mu p^\nu}$ is not gauge invariant (under \eqref{gauge-t} $m\rightarrow\Omega^{-1}m$.) Yet one may investigate the ratio of masses which is indeed a gauge invariant quantity.


Let us illustrate with the help of a thought experiment, why path-dependent mass variation under parallel transport \eqref{master-eq-mass}, represents a serious phenomenological issue. Let us assume that at the origin ($x=0$), we have a collection of $n$ (identical) hydrogen atoms, each with an associated electron of mass: $m_0\equiv m_e(0)$. Let us further assume that each individual atom is parallel transported from the origin to a spacetime point $x$, following different paths in $W_4$ spacetime. The trajectory $\cal C$, joining the origin and the point $x$ for each atom, can be seen as a small deviation of the trajectory of any other hydrogen atom in the collection: ${\cal C}_{i+1}={\cal C}_i+\delta{\cal C}$. Upon arriving at $x$ the mass of the electron associated to the atom labeled $j$, according to \eqref{master-eq-mass} reads:

\bea m_j(x)=m_0\,e^{\int_{{\cal C}_j}\left(Q^*_\lambda+\frac{1}{2}Q_{\lambda\mu\nu}u^\mu u^\nu\right)dx^\lambda},\label{transp-h2'}\eea while in $\tilde W_4$, where vectorial nonmetricity takes place \eqref{vect-nm}, we have:

\bea m_j(x)=m_0\,e^{\frac{1}{2}\int_{{\cal C}_j}Q_\mu dx^\mu}.\label{transp-h2}\eea 

After parallel transport, assuming that the field $Q_{\alpha\mu\nu}$ is not vanishing along the trajectory, we get that at point $x$ the electron associated to each atom has a slightly different mass compared with the remaining ones in the collection of atoms. This means that, on top of the standard red(blue)shift of spectral lines taking place in Riemann $V_4$ spacetimes, the spectral lines in the emission (or absorption) spectrum of each atom, suffer a slight broadening. This is to be contrasted with the sharpness of the spectral lines in the hydrogen atom. The argument, very similar to the one stated by Einstein and leading to rejection of the original Weyl theory \cite{weyl-1917}, is applicable to any other atom, or collection of atoms, including the caesium-133 atom, on which the modern atomic clocks are based.


\subsection{Qualitative and quantitative estimates of the SCE in the coincident gauge of $Z_4$}\label{subsect-sce-z4}


In generalized Weyl spacetime $W_4$, which includes the teleparallel $Z_4$ space as a particular case, masses of different particles vary in different ways as these follow different paths in spacetime \eqref{transp-h2'}. This effect is different from the loss of synchronization that arises in general relativity and is related to the so called second clock effect \cite{adler-book, perlick-1991, novello-1992, tucker, 2clock-1, 2clock-2, 2clock-3, 2clock-4, 2clock-5, 2clock-delhom, 2clock-no-1, 2clock-hobson, 2clock-tomi, tomi-replay}. Let us explore how this effect arises in the coincident gauge of teleparallel spacetime $Z_4$, where we consider a static and spherically symmetric metric \eqref{sp-symm-met}. Let us consider a collection of identical atoms that are parallel transported along neighboring paths from the origin $x=0$ to a given point $x$, following the same radial trajectory: ${\cal C}=\{(r,\theta,\vphi)|R\leq r\leq R+h,\theta=\pi/2,\vphi=\vphi_0\}$. This amounts to great simplification of calculations since only the dependence on the four-velocity $u^\alpha:=dx^\alpha/d\tau$, matters. The mass of any atom in the collection varies according to \eqref{master-eq-mass}:

\bea m(x)=m_0e^Q\,e^{\frac{1}{2}\int_{\cal C}Q_{\lambda\mu\nu}u^\mu u^\nu dx^\lambda},\label{2clock-z4-1}\eea where the quantity $Q=\int_{\cal C}Q^*_\lambda dx^\lambda$ is the same for any atom in the collection since the trajectory of parallel transport is the same for all of them. Let us further assume, for definiteness, that the larger speed of parallel transport is attained by certain atom labeled 1: $dr/d\tau=\alpha$ -- a constant, while the smaller speed is for atom labeled 2: $dr/d\tau=\beta$ -- another constant, such that $\alpha\neq\beta$. Let us take the larger difference arising between the masses of any two atoms in the collection at $x$: $\Delta m=m_2(x)-m_1(x)$, where

\bea m(x)=m_0e^Q\,e^{\frac{1}{2}\int_R^{R+h}\left[Q_{rrr}\left(\frac{dr}{d\tau}\right)^2+Q_{r00}\left(\frac{dt}{d\tau}\right)^2\right]dr},\label{2clock-z4-2}\eea and, since we are working with the static, spherically symmetric background metric \eqref{sp-symm-met}, for radial motion we have that

\bea 1=A^2\left(\frac{dt}{d\tau}\right)^2-B^2\left(\frac{dr}{d\tau}\right)^2.\label{2clock-z4-3}\eea Hence, equation \eqref{2clock-z4-2} transforms into:

\bea m(x)=m_0e^{Q+\ln\frac{A(R+h)}{A(R)}}e^{\int_R^{R+h}\left(\frac{A_r}{A}-\frac{B_r}{B}\right)B^2\left(\frac{dr}{d\tau}\right)^2dr}.\label{2clock-z4-4}\eea In order to go further with our computations we have to make additional assumptions on the metric. Otherwise, if invoke a specific theory of gravity, we do not need additional assumptions in order to get qualitative results. Let us assume that Schwarzschild metric is a solution of the equations of motion of a given gravitational theory. Then:

\bea &&A^2=1-\frac{2GM}{r},\;B=A^{-1},\nonumber\\
&&\frac{B_r}{B}=-\frac{A_r}{A},\;\frac{A_r}{A}=\frac{GM}{A^2r^2},\label{add-assumpt}\eea where $G$ is Newton's constant and $M$ is the mass of the central body. We further assume that $h\ll R$ (the height of the distant point is much less that the radius of the central body), so that $h/R$ is a small parameter. Then, from \eqref{2clock-z4-4} one gets the following gauge invariant ratio: 

\bea \frac{\Delta\omega_{if}}{\omega_{if}}=\frac{\Delta m}{m}=1-e^{\frac{2(\alpha^2-\beta^2)GMh}{(R-2GM)^2}},\label{mass-ratio}\eea where $\Delta\omega_{if}$ quantifies the broadening of the given spectral line. Omitting terms ${\cal O}(h^2/R^2)$ and higher, from \eqref{mass-ratio} we obtain:

\bea \frac{|\Delta\omega_{if}|}{\omega_{if}}\approx\frac{2|\alpha^2-\beta^2|GM/R}{(1-2GM/R)^2}\left(\frac{h}{R}\right).\label{mass-var}\eea It is seen from this equation that the shift of frequencies of individual atoms is proportional to the frequency: $\Delta\omega_{if}\propto\omega_{if}$. This is in contrast to GR redshift of frequencies which is the same for every frequency. 

The dependence on the speed in \eqref{mass-var} is manifest through the constants $\alpha$ and $\beta$. Although the SCE may be very tiny for astrophysical objects like planets and normal stars, for compact objects it my become significant. In order to make an estimate of the effect, let us choose a typical neutron star of $2$ Solar masses and radius $R\approx 12$ km. Let us further arbitrarily set the constant speeds $\alpha\sim 10^{-1}$ and $\beta\sim 10^{-2}$, respectively. This amounts to having a speed $\alpha$ of the order of the scape velocity from the surface of the neutron star (recall that we are using the units where the speed of light $c=1$.) In this case, since $GM/R\approx 0.25$, our estimate reads: $$\frac{|\Delta\omega_{if}|}{\omega_{if}}\approx 2\times 10^{-2}\frac{h}{R},$$ or if set $h<10^{-2}R$, which amounts to $h<100$ m, we get that $|\Delta\omega_{if}|/\omega_{if}<2\times 10^{-4}$. Although the effect of mass variation is significantly stronger for most compact objects, for other astrophysical objects it must be measurable as well. Notice that the strength of the effect may considerably increase if consider some neighborhood of the event horizon of a Schwarzschild black hole. 

We want to underline that expression \eqref{mass-var} and consequent estimates, highly depend on the validity of Schwarzschild solution, i. e., on a specific theory of gravity. Yet, the effect of mass variation itself is independent of the postulated theory or metric.


\section{Perihelion shift in $Z_4$}\label{sect-tests}


Let us to give estimates for the classic tests of gravity, in particular for the perihelion shift and for the light bending. As above we assume the static, spherically symmetric metric \eqref{sp-symm-met} in the coincident gauge of teleparallel $Z_4$ space. For the planar motion where $\theta=\pi/2$ (this will be the case of interest below) the non-vanishing components of the non-metricity are:

\bea &&Q_{r00}=2AA_r,\;Q_{rrr}=-2BB_r,\nonumber\\
&&Q_{r\theta\theta}=Q_{r\vphi\vphi}=-2r,\label{non-0-qanm}\eea while the only non-vanishing component of the vector $Q_\alpha=g^{\kappa\lambda}Q_{\kappa\lambda\alpha}$ is: 

\bea Q_r=-2\frac{B_r}{B}\;\Rightarrow\;Q=\int Q_\mu dx^\mu=-2\ln B.\label{non-0-qa}\eea

In this setup the geodesic equations \eqref{geod-w4-1} for a point particle with non-vanishing mass and \eqref{0-geod-w4} for massless particles, read:

\bea &&\frac{d^2x^\alpha}{ds^2}-\left(Q^{\;\;\alpha}_{\mu\;\;\nu}-\frac{1}{2}Q^\alpha_{\;\;\mu\nu}\right)\frac{dx^\mu}{ds}\frac{dx^\nu}{ds}\nonumber\\
&&\;\;\;\;\;\;\;\;\;\;\;=\left(Q^*_\mu-\frac{1}{2}Q_{\mu\kappa\lambda}\frac{dx^\kappa}{ds}\frac{dx^\lambda}{ds}\right)h^{\mu\alpha},\label{geod-z4}\\
&&\frac{dk^\alpha}{d\lambda}+\{^\alpha_{\mu\nu}\}k^\mu k^\nu=0,\label{0-geod-z4}\eea respectively, where the orthogonal projection tensor $h^{\alpha\beta}$ is defined in \eqref{proj-t}. 

The fact that the null-geodesics in $W_4$ \eqref{0-geod-z4} coincide with the Riemannian geodesics means that photons and radiation interact only with the LC curvature of the spacetime. Hence, the bending of light in $W_4$ will be the same as the one predicted by GR in $V_4$.

In the remainder of this section, in order to simplify computations and without loss of generality, in the definition of $Q^*_\alpha$ \eqref{qaster}, we shall set the constant $b=0$, so that $Q^*_\alpha=Q_\alpha=Q^\mu_{\;\;\alpha\mu}=Q^\mu_{\;\;\mu\alpha}$.


\subsubsection{Planar motion in teleparallel $Z_4$ spacetime}


Given that below we shall focus in planar motion in static, spherically symmetric spacetime in the coincident gauge of $Z_4$, it is necessary to show that planar motion is described by the equations of motion in this setup. In what follows we consider the approximation where terms ${\cal O}(v^3)$ and higher ($v\sim dr/ds\sim rd\vphi/ds\sim rd\theta/ds$), are omitted from the computations. In this approximation, for instance: 

\bea \frac{1}{2}Q_{\lambda\mu\nu}\frac{dx^\lambda}{ds}\frac{dx^\mu}{ds}\frac{dx^\nu}{ds}\approx-\frac{Q_{r00}}{2A^2}\frac{dr}{ds}.\label{aprox}\eea 

The $\theta$-component of the motion equation \eqref{geod-z4} for a point particle with non-vanishing mass in the coincident gauge reads:

\bea &&\frac{d^2\theta}{ds^2}=-\frac{2}{r}\frac{dr}{ds}\frac{d\theta}{ds}+\frac{d}{ds}\left(\ln\frac{B^2}{A}\right)\frac{d\theta}{ds}\nonumber\\
&&\;\;\;\;\;\;\;\;\;\;\;\;+2\sin\theta\cos\theta\left(\frac{d\vphi}{ds}\right)^2,\nonumber\eea or

\bea &&\frac{d}{ds}\left(r^2\frac{d\theta}{ds}\right)=r^2\frac{d}{ds}\left(\ln\frac{B^2}{A}\right)\frac{d\theta}{ds}\nonumber\\
&&\;\;\;\;\;\;\;\;\;\;\;\;\;\;\;\;\;\;\;\;\;\;+2r^2\sin\theta\cos\theta\left(\frac{d\vphi}{ds}\right)^2.\label{plan-geod-z4}\eea If make the point-dependent replacement of affine parameter, $ds=Ad\bar s/B^2$, the above equation can be rewritten in the following way:

\bea \frac{d}{d\bar s}\left(r^2\frac{d\theta}{d\bar s}\right)=2r^2\sin\theta\cos\theta\left(\frac{d\vphi}{d\bar s}\right)^2,\label{plan-redef}\eea where the only difference with the corresponding component of the geodesic equation of Riemann geometry spacetimes is in the factor $2$ in the right-hand side. Hence, if at some initial time $\theta=\pi/2$ (equatorial plane) and $d\theta/d\bar s=0$, equation \eqref{plan-redef} predicts that $\theta=\pi/2$ for all times. This result is obviously valid as well in terms of the original affine parameter $s$. Hence, in our setup -- coincident gauge of teleparallel $Z_4$ static, spherically symmetric space -- planar motion is predicted by the equations of motion. Below we shall consider, precisely, motions in the $\theta=\pi/2$ plane.


\subsubsection{Perihelion shift}


Let us study motions in the equatorial plane, where $\theta=\pi/2$, $d\theta/ds=d\theta/d\lambda=0$. In this specific case, for particles with mass we have that:

\bea 1=-A^2\left(\frac{dt}{ds}\right)^2+B^2\left(\frac{dr}{ds}\right)^2+r^2\left(\frac{d\vphi}{ds}\right)^2.\label{metric-eq}\eea  

For static, spherically symmetric metric in the coincident gauge of teleparallel $Z_4$ spacetime, the equations of motion \eqref{geod-z4} for particles with mass read:

\bea &&\frac{d^2t}{ds^2}=\frac{d}{ds}\left(\ln\frac{B^2}{A^3}\right)\frac{dt}{ds}\;\Rightarrow\;A^2\frac{dt}{ds}=\frac{\alpha_0B^2}{A},\nonumber\\
&&\frac{d^2\vphi}{ds^2}=\frac{d}{ds}\left(\ln\frac{B^2}{r^2A}\right)\frac{d\vphi}{ds}\;\Rightarrow\;r^2\frac{d\vphi}{ds}=\frac{\beta_0B^2}{A},\label{geod-z4-sss}\eea where $\alpha_0$ and $\beta _0$ are integration constants and, as before, we have omitted terms ${\cal O}(v^3)$ and higher. 

Let us stress that, in order to be able to compare with experimental results, just as in standard calculations in the literature, we have to replace $ds$ everywhere by $id\tau$, where $\tau$ is the proper time and $i$ is the imaginary unit. Simultaneously we have to replace $\alpha_0\rightarrow-ik$ and $\beta_0\rightarrow-ih$. Hence, we may rewrite \eqref{geod-z4-sss} in the following way ($k$ and $h$ are constants):

\bea A^2\frac{dt}{d\tau}=k\frac{B^2}{A},\;r^2\frac{d\vphi}{d\tau}=h\frac{B^2}{A},\label{geod-dtau}\eea while \eqref{metric-eq} now reads:

\bea 1=A^2\left(\frac{dt}{d\tau}\right)^2-B^2\left(\frac{dr}{d\tau}\right)^2-r^2\left(\frac{d\vphi}{d\tau}\right)^2.\label{metric-tau}\eea Substituting equations \eqref{geod-dtau} in \eqref{metric-tau} we get that:

\bea B^2\left(\frac{dr}{d\tau}\right)^2=-1+\frac{k^2B^4}{A^4}-\frac{h^2B^4}{A^2r^2},\label{geod-z4-req}\eea which, combined with second equation in \eqref{geod-dtau}, yields:

\bea \frac{B^2}{r^2}\left(\frac{dr}{d\vphi}\right)^2=-1+\left(\frac{k^2}{A^2}-\frac{A^2}{B^4}\right)\frac{r^2}{h^2}.\label{vphi-geod-z4}\eea This equation is to be contrasted its general relativistic equivalent (the expression can be found in any textbook):

\bea \frac{B^2}{r^2}\left(\frac{dr}{d\vphi}\right)^2=-1+\left(\frac{k^2}{A^2}-1\right)\frac{r^2}{h^2}.\label{vphi-geod-gr}\eea 

In order to be able to make estimates let us choose the static, spherically symmetric Schwarzschild metric:

\bea AB=1,\;A^2=1-\frac{2m}{r},\label{schw-met}\eea where $m\equiv GM$, with $G$ -- Newton's constant, and $M$ -- the mass of the central body. Besides, we introduce a new appropriate variable: $u\equiv 2m/r$. Since we consider the case where $r\ll 2m$, then $u$ is a small quantity of first order of smallness $\sim{\cal O}(1)$ like the quantity $m^2/h^2\sim{\cal O}(1)$. Omitting terms $\sim{\cal O}(4)$ and higher, equation \eqref{vphi-geod-z4} can be rewritten in the following way:

\bea &&(u')^2=\frac{4m^2}{h^2}(k^2-1)+\frac{16m^2}{h^2}u\nonumber\\
&&\;\;\;\;\;\;\;\;\;\;\;-\left(1+\frac{24m^2}{h^2}\right)u^2+u^3,\label{geod-u-z4}\eea where the prime denotes derivative with respect to the variable $\vphi$. Let us further derive \eqref{geod-u-z4}:

\bea &&u''+u=\bar a+\frac{3\bar\alpha}{2}u^2,\nonumber\\
&&\bar a=\frac{8m^2}{h^2}\bar\alpha,\;\bar\alpha=\left(1-\frac{24m^2}{h^2}\right),\label{2-der-eq}\eea where we have made the replacement $\vphi\rightarrow\vphi/\sqrt{1+24m^2/h^2}$. If we compare \eqref{2-der-eq} with the similar equation taking place in the framework of general relativity:

\bea u''+u=a_\text{GR}+\frac{3}{2}u^2,\label{2-der-gr}\eea where we have defined the constant $a_\text{GR}\equiv 2m^2/h_\text{GR}^2$, we see that there is a difference in the quadratic term $\sim u^2$, which is the one that generates the non-periodicity of orbital motion. As we shall see, this will have consequences for the predicted perihelion shift.\footnote{We want to emphasize that, despite that $a_\text{GR}$ in \eqref{2-der-gr} and $\bar a$ in \eqref{2-der-eq} do not coincide, this has not observational consequences since what we compare are the terms in the mentioned equations with the corresponding terms in the equation of the ellipse laying on the plane $\theta=\pi/2$.} 

Following a standard procedure we can find the solution of \eqref{2-der-eq} in the form of the following sum:

\bea u=u_1+u_2,\label{to-sol-1}\eea where $u_1\sim{\cal O}(1)$ while $u_2\sim{\cal O}(2)$. We omitt terms $\sim{\cal O}(3)$ and higher. Substituting \eqref{to-sol-1} into \eqref{2-der-eq} we get two differential equations:

\bea &&u''_1+u_1=\bar a,\nonumber\\
&&u''_2+u_2=\frac{3\bar\alpha}{2}u^2_1.\label{to-sol-2}\eea The solution of the first equation above can be found in the form:

\bea u_1=\bar a+\bar A\cos\vphi,\label{to-sol-u1}\eea where the amplitude $\bar A$ is an integration constant. Substituting this into the right-hand side of the second equation in \eqref{to-sol-2}, and considering terms up to $\sim{\cal O}(2)$, one gets:

\bea u''_2+u_2=\frac{3\bar\alpha}{2}\left(\bar a^2+\frac{\bar A^2}{2}\right)+3\bar\alpha\bar a\bar A\cos\vphi+3\bar\alpha\bar A^2\cos 2\vphi.\nonumber\eea The solution is given by:

\bea u_2=\frac{3\bar\alpha}{2}\left(\bar a^2+\frac{\bar A^2}{2}\right)+\frac{3\bar\alpha}{2}\bar a\bar A\vphi\sin\vphi-\bar\alpha\bar A^2\cos 2\vphi.\nonumber\eea The second term above ($\propto\vphi\sin\vphi$) is the one responsible for the perihelion shift since it breaks the periodicity of the orbital motion. Notice that, although this term is $\sim{\cal O}(2)$ its importance increases as the orbital motion proceeds, since $\vphi$ increases. In the reminder of this subsection we shall omit constant and periodic terms of order $\sim{\cal O}(2)$ and higher, yet the term $\propto\vphi\sin\vphi$ can not be omitted for the mentioned reason. In this approximation the full solution of \eqref{2-der-eq} is found to be:

\bea u=\bar a+\bar A\cos\vphi+\frac{3\bar\alpha}{2}\bar a\bar A\vphi\sin\vphi,\nonumber\eea or, since $\cos(\vphi-3\bar\alpha\bar a\vphi/2)=\cos\vphi+3\bar\alpha\bar a\vphi\sin\vphi/2$:

\bea u=\bar a+\bar A\cos\left(\vphi-\frac{3\bar\alpha}{2}\bar a\vphi\right).\label{full-u-sol}\eea This is the equation of an ellipse laying on the plane $\theta=\pi/2$ ($R$ is its semi-major axis and $e$ is its eccentricity):

\bea &&\frac{1}{r(\vphi)}=\frac{\bar a}{2m}+\frac{\bar A}{2m}\cos\left(\vphi-\frac{3\bar\alpha}{2}\bar a\vphi\right)\nonumber\\
&&\;\;\;\;\;\;\;\;\;=\frac{1}{l}+\frac{e}{l}\cos(\vphi-\vphi_0),\label{ellipse}\eea where $l=R(1-e^2)$ is the so called semi-latus rectum of the ellipse and $\vphi_0$ is the orientation of the principal axis of the ellipse. Comparing the parameters of the ellipse allows to identify the constants of the theory: 

\bea \bar a=\frac{2m}{l}\approx\frac{8m^2}{h^2}\;\Rightarrow\;h^2=4ml,\;\bar A=\frac{2me}{l}.\label{phys-par-z4}\eea A similar comparison in the GR framework yields:

\bea a_\text{GR}=\frac{2m^2}{h^2_\text{GR}}=\frac{2m}{l}\;\Rightarrow\;h^2_\text{GR}=ml,\;A_\text{GR}=\frac{2me}{l}.\label{phys-par-gr}\eea From \eqref{ellipse} the following perihelion shift amount may be inferred:

\bea \delta\vphi=3\pi\bar\alpha\bar a=3\pi\bar\alpha a_\text{GR}=\bar\alpha\delta\vphi_\text{GR},\label{perih-z4}\eea where $\delta\vphi_\text{GR}=6\pi m/l$ is the magnitude of the perihelion shift predicted by GR theory. Hence, the relative difference between the effect predicted by the present geometric setup (coincident gauge of teleparallel $Z_4$ geometry) and the one predicted by general relativity amounts to:

\bea &&\frac{\delta\vphi_\text{GR}-\delta\vphi}{\delta\vphi_\text{GR}}=\frac{24m^2}{h^2}=\frac{6m}{l}=\frac{\delta\vphi_\text{GR}}{\pi},\nonumber\\
&&\delta\vphi=\delta\vphi_\text{GR}-\frac{\left(\delta\vphi_\text{GR}\right)^2}{\pi}.\label{perih-rel-diff}\eea I. e., it represents an $\sim{\cal O}(2)$ contribution to the GR effect.


\section{Challenges of the SCE}\label{sect-sce-challenge}


Several authors have argued that there is a loophole in the line of reasoning leading to rejection of Weyl geometry, being a phenomenologically non-viable scenario for physical theories, since classical physics does not describe atomic phenomena without quantum theoretical modifications (see, for instance the argument in the paragraph at the beginning of page 506 of \cite{adler-book}.) However, it has been discussed in the bibliography the potential of Weyl geometry to describe quantum-mechanical phenomena \cite{novello-1992, smolin-1979, cheng-1988, cheng-arxiv, london-1927, adler-1970, ross-1976, godfrey-1984, wheeler-1990, castro-1991, wood-1997, arabs-2003, novello-2011, italia-2012}. Besides, although the atomic spectrum is of quantum origin, the mass of the electron, as any other mass, is a classical quantity, subject to the classical laws of gravity. 

Other authors have (apparently) demonstrated that the SCE does not arise neither in $\tilde W_4$ \cite{2clock-no-1} nor in $Z_4$ \cite{2clock-tomi} spaces. But when one looks into the corresponding demonstrations one notices that the relevant expressions, for instance the one driving the change of the inner product of vectors along a given curve, are not gauge covariant equations. Thus the most salient feature of generalized Weyl spaces, including standard Weyl $\tilde W_4$ and teleparallel geometric $Z_4$ setups: gauge symmetry, is out of consideration. Let us to discuss in more detail the most recent arguments raised against the SCE in references \cite{2clock-no-1} and in \cite{2clock-tomi}.


\subsection{Recent arguments against the SCE}\label{subsect-sce}


As stated in section \ref{sect-2clock}, in reference \cite{2clock-no-1} a point of view is exposed according to which the SCE does not take place in standard Weyl space $\tilde W_4$ with vectorial nonmetricity \eqref{vect-nm}. The argument in \cite{2clock-no-1} is based in equation (49) of that paper, which describes the evolution of the inner product of two given vectors ${\bf v}$ and ${\bf w}$, which are parallel transported along some curve ${\cal C}$:

\bea \frac{d}{d\lambda}({\bf v},{\bf w})=\frac{d}{d\lambda}\left[g_{\mu\nu}v^\mu w^\mu\right]=0,\label{49}\eea where $\lambda$ is a parameter along ${\cal C}$ and $({\bf v},{\bf w}):=g_{\mu\nu}v^\mu w^\mu$. But the above equation -- equation (49) of reference \cite{2clock-no-1} -- shares the same problem with equations \eqref{length-parallel-t} and \eqref{length-change} of the present paper. The problem is that, in general (arbitrary conformal weight) it is not gauge covariant. Actually, let us assume that $w({\bf v})=k$ while $w({\bf w})=l$, are the conformal weights of vectors ${\bf v}$ and ${\bf w}$, respectively. Hence, the conformal weight of their inner product: $w({\bf v},{\bf w})=2+k+l$. In consequence, under the gauge transformations \eqref{gauge-t},

\bea \frac{d}{d\lambda}({\bf v},{\bf w})=0\rightarrow\frac{d}{d\lambda}({\bf v},{\bf w})=-(2+k+l)\frac{d\ln\Omega}{d\lambda}({\bf v},{\bf w}).\nonumber\eea As seen the condition $d({\bf v},{\bf w})/d\lambda=0$ is gauge covariant only when $k+l=-2$. As an example, consider the particular case when ${\bf v}={\bf w}$ is a tangent vector with weight $w=-1$. In this case, $({\bf v},{\bf v})=v^2$, while $l=k=-1$, so that, under \eqref{gauge-t}: $$\frac{dv^2}{d\lambda}=0\rightarrow\frac{dv^2}{d\lambda}=0,$$ i. e., the above equation is gauge covariant. But there are other vectors such as, for instance, the four-momentum ${\bf p}$, with components $p^\mu=mdx^\mu/d\tau$, or the wave vector ${\bf k}$, whose conformal weight is $w({\bf p})=w({\bf k})=-2$. In this case the gauge transformation of the parallel transport law \eqref{49} amounts to ($p^2=g_{\mu\nu}p^\mu p^\nu$):

\bea \frac{dp^2}{d\lambda}=0\rightarrow\frac{dp^2}{d\lambda}=2\frac{d\ln\Omega}{d\lambda}p^2,\nonumber\eea so that it is not gauge covariant.\footnote{For the wave vector, since $k\equiv||{\bf k}||=0$, gauge symmetry is trivially preserved by \eqref{49}.} This is contrary to the spirit of the reasoning line in \cite{2clock-no-1} where it is adopted the natural (gauge) covariant derivative identified in the geometric interpretation of Weyl gauge theories \cite{dirac-1973}. Hence, the counterargument exposed in that reference fails to give a correct explanation of non occurrence of the SCE. The correct analysis requires to replace \eqref{49} -- equation (49) in \cite{2clock-no-1} -- by equation \eqref{master-eq} of the present paper. This means that, under the assumption of the parallel transport law \eqref{ptp} and of the consistency hypothesis \eqref{mass-law'}, the second clock effect takes place in $W_4$ (also in its particular subclass $\tilde W_4$).


In lemma 2 of reference \cite{2clock-tomi} it is shown that in teleparallel $Z_4$ (and other teleparallel) spaces, the SCE does not arise. The demonstration of lemma 2 is based in the following equation -- equation (12) of the mentioned reference -- that drives the change of the inner product of vectors ${\bf v}$ and ${\bf w}$ when parallel transported along the closed curve ${\cal C}_\gamma$:\footnote{Notice that our signature for the nonmetricity is contrary to that of \cite{2clock-tomi}. Actually, in this reference $\nabla_\alpha g_{\mu\nu}=Q_{\alpha\mu\nu}$, so that one has to make the replacement $Q_{\alpha\mu\nu}\rightarrow-Q_{\alpha\mu\nu}$ in order to meet our conventions.}

\bea \Delta({\bf v},{\bf w})=-\oint_{{\cal C}_\gamma}Q_{\lambda\mu\nu}v^\mu w^\nu dx^\lambda,\label{12}\eea where $\Delta({\bf v},{\bf w})$ means the total change of the inner product. This is the equivalent of our equation \eqref{length-parallel-t} which is not gauge invariant. Hence, the demonstration of lemma 2 of \cite{2clock-tomi} is valid only if ignore Weyl gauge symmetry. A gauge invariant version of their demonstration, that should be based on our equation \eqref{master-int} with replacement $\int_{\cal C}\rightarrow\oint_{{\cal C}_\gamma}$, can be found in \cite{quiros-arxiv-2022}. It is demonstrated that if consider gauge symmetry the lemma 2 of reference \cite{2clock-tomi} is incorrect. Besides, as shown in \cite{quiros-arxiv-2022}, there are other inconsistencies in the demonstration of lemma 2.


\section{Relevance of the underlying postulates}\label{sect-postu}


One can find in the bibliography, absolutist approaches to the SCE like the one exposed in the following statement in reference \cite{tomi-replay}: ``the second clock effects have nothing to do with geometry but are entirely determined by the matter couplings...'' Similar absolutist approaches to the inevitability of the SCE are found in the bibliography as well. Here we shall show that absolutism is not of help in discussions about the SCE. In this regard, we shall point out the relevance of postulates on the occurrence (or not) of the second clock effect and why the answer to the question about the occurrence of this effect has no absolute or correct answer, unless these postulates are clearly stated.

In the present paper, for instance, all of our statements and results are based in the parallel transport postulate \eqref{ptp} and in the consistency hypothesis stated in section \ref{sect-geod}, more specifically in subsection \ref{subsect-consist-hypo}. While the parallel transport postulate is necessary for the definition of vectorial and/or tensorial operations, as well as of differential operations in $W_4$, the consistency hypothesis allows to identify hypothetical vectors and tensors living in $W_4$, with the related physical vectors and tensors. For instance, if we identify the four-momentum ${\bf p}$ of a (spinless) test particle with the related four-momentum \eqref{4-p} in $W_4$, then, according to \eqref{master-int-vec}:

\bea p(x)=p(0)\;\exp{\left[\int_{\cal C}\left(Q^*_\alpha-\frac{1}{2}Q_{\alpha\mu\nu}t^\mu_{\bf u}t^\nu_{\bf u}\right)dx^\alpha\right]}.\label{pi-eq}\eea Due to the above identification, we must identify $p\equiv im$, where $m$ is the mass of the test particle. Hence, equations \eqref{master-eq-mass} and \eqref{mass-law'} arise as a consequence. This identification can be only the result of a hypothesis or postulate, that links a physical phenomenon with an assumed geometrical background (in the present case it is $W_4$ space).

Renouncing to the above consistency hypothesis \eqref{mass-law'} amounts to accepting that equations \eqref{master-eq-mass} and \eqref{pi-eq} are not valid, so that one must assume another hypothesis on the nature of the mass of point particles, for instance, one that could led to avoidance of the SCE. Otherwise the nature of the mass $m$ remains undetermined. 

Let us explore the consequences of assuming different parallel transport laws and of making, at the same time, different specific choices of the term $\propto\delta m/\delta x^\alpha$ in the timelike geodesic equation \eqref{geod-w4}:

\bea \frac{d^2x^\alpha}{ds^2}+\{^\alpha_{\mu\nu}\}\frac{dx^\mu}{ds}\frac{dx^\nu}{ds}-\frac{\delta\ln m}{\delta x^\mu}h^{\mu\alpha}=0.\nonumber\eea 

There are several hypotheses or postulates one can make about this term but bellow, for simplicity, we shall focus in three different examples that serve as illustrations of the kind of discussion we want to hold.


\subsection{Example 1}\label{example-1}


For instance, one may postulate that 

\bea m=m_0=\text{const.}\;\Rightarrow\;\frac{\delta m}{\delta x^\alpha}=0.\label{hipo-1}\eea In this case timelike (spinless) test (point) particles follow geodesics of Riemann geometry. Hence, gauge invariance is not a manifest symmetry of the equations of motion so that we may dispense with this symmetry. 

Let us further assume an appropriate parallel transport law along the worldline $x^\alpha(\xi)$ (compare with \eqref{ptp}):

\bea \frac{D{\bf T}}{d\xi}=\frac{dx^\mu}{d\xi}\nabla_\mu{\bf T}=0,\label{ptp-1}\eea where ${\bf T}$ is an arbitrary tensor as it has been previously defined in section \ref{sect-gaug-der}.

The four-momentum of test particles ${\bf p}=m_0{\bf u}$ obeys that $g_{\mu\nu}p^\mu p^\nu=-m_0^2$. If take the covariant derivative of the latter equation along given path $x^\alpha(\xi)$, one gets:

\bea \frac{Dg_{\mu\nu}}{d\xi}p^\mu p^\nu+2g_{\mu\nu}p^\nu\frac{Dp^\mu}{d\xi}=0,\nonumber\eea but since, according to \eqref{ptp-1}, $Dp^\alpha/d\xi=0$, then, from the above equation it follows that 

\bea \nabla_\alpha g_{\mu\nu}=0\;\Rightarrow\;Q_{\alpha\mu\nu}=0.\label{res-1}\eea Hence, assumption of the hypothesis that $\delta m/\delta x^\alpha=0,$ together with the parallel transport law \eqref{ptp-1}, single out Riemann geometry as the geometric setup of background space in this case ($m=m_0$).


\subsection{Example 2}\label{example-2}


Let us choose the following postulate on the mass parameter \cite{2clock-hobson}: 

\bea m=m_0\phi\;\Rightarrow\;\frac{\delta\ln m}{\delta x^\alpha}=\der_\alpha\ln\phi,\label{hipo-2}\eea where $\phi$ is a scalar field. For simplicity lets assume the parallel transport law \eqref{ptp-1}. The covariant derivative of equation $-m^2=g_{\mu\nu}p^\mu p^\nu$, along the worldline $x^\alpha(\xi)$, leads to:

\bea -2m\frac{dm}{d\xi}=m^2\frac{Dg_{\mu\nu}}{d\xi}u^\mu u^\nu+2mg_{\mu\nu}u^\nu\frac{Dp^\mu}{d\xi}.\label{ex-21}\eea Taking into account the parallel transport law \eqref{ptp-1}: $Dp^\alpha/d\xi=0$, from the above equation it follows that

\bea \frac{d\ln m}{d\xi}=-\frac{1}{2}\frac{dx^\lambda}{d\xi}\nabla_\lambda g_{\mu\nu}u^\mu u^\nu,\label{ex-22}\eea or, equivalently

\bea \frac{dx^\lambda}{d\xi}\left(\der_\lambda\ln m+\frac{1}{2}\nabla_\lambda g_{\mu\nu}u^\mu u^\nu\right)=0,\label{ex-23}\eea from where it follows that $\der_\alpha\ln\phi=Q_{\alpha\mu\nu}u^\mu u^\nu/2,$ or 

\bea Q_{\alpha\mu\nu}=-2\der_\alpha\ln\phi g_{\mu\nu}.\label{res-2}\eea This nonmetricity corresponds to WIG space, which is a subset of standard Weyl space $\tilde W_4$ when the gauge vector is the gradient of a scalar field: $Q_\alpha=-2\der_\alpha\ln\phi$. In this case the scalar field participates in the definition of the nonmetricity of spacetime, so it is a scalar field of geometric nature. 

Alternatively, if assume that the scalar field is a non-geometric (compensator) field, as in \cite{2clock-hobson}, then one can not identify the physical four-momentum of the test particle with the hypothetical four-momentum in $W_4$ whose length obeys \eqref{master-eq-mass}. In other words: equations \eqref{ex-21}-\eqref{res-2} and \eqref{master-eq-mass}, do not take place. Hence, we can not describe the motion of particles with the mass obeying \eqref{hipo-2} and where $\phi$ is a non-geometric scalar field, in generalized Weyl space $W_4$. A potentially feasible description in this case could be that a fifth force $f^\alpha_5$ arises in Riemann space $V_4$, such that

\bea \frac{Dp^\alpha}{ds}=f^\alpha_5=-m_0g^{\alpha\mu}\der_\mu\phi.\label{ex-2-alter}\eea


\subsection{Example 3}\label{example-3}


Finally, les us assume that equation \eqref{hipo-2} takes place, together with the parallel transport law \eqref{ptp}:

\bea \frac{D^*{\bf T}}{d\xi}:=\frac{dx^\mu}{d\xi}\nabla^*_\mu{\bf T}=0.\nonumber\eea This means that we are considering Weyl gauge symmetry as a manifest symmetry of generalized Weyl spaces (this is precisely the case considered in \cite{2clock-hobson}). We have that (recall that, due to \eqref{ptp}, $D^*p^\alpha/d\xi=0$):

\bea &&-\frac{D^*m^2}{d\xi}=\frac{dx^\lambda}{d\xi}\nabla^*_\lambda g_{\mu\nu}p^\mu p^\nu\nonumber\\
&&\;\;\;\;\;\;\Rightarrow\;\frac{d\ln m}{d\xi}=\frac{dx^\lambda}{d\xi}\left(Q^*_\lambda+\frac{1}{2}Q_{\lambda\mu\nu}u^\mu u^\nu\right)\nonumber\\
&&\;\;\;\;\;\;\Rightarrow\;\der_\alpha\ln\phi-\left(Q^*_\alpha+\frac{1}{2}Q_{\alpha\mu\nu}u^\mu u^\nu\right)=0.\label{ex-31}\eea From the last equation it follows that: 

\bea Q_{\alpha\mu\nu}=2\left(Q^*_\alpha-\der_\alpha\ln\phi\right)g_{\mu\nu}.\label{ex-32}\eea This case corresponds to vectorial nonmetricity with gauge vector $Q_\alpha=2\left(Q^*_\alpha-\der_\alpha\ln\phi\right)$, i. e., it corresponds to standard Weyl space $\tilde W_4$. 

The above analysis is correct only if the scalar field $\phi$ is a gauge field of geometrical nature which determines the nonmetricity of space. Alternatively, if consider that $\phi$ is a compensator field of non-geometric origin, then equations \eqref{ex-31}, \eqref{ex-32} are not satisfied. Means that the physical four-momentum of spinless test particles can not be identified with the corresponding hypothetical four-momentum in $W_4$ space. In this last case we must renounce to describing the motion of physical test bodies in generalized Weyl space $W_4$.


\subsection{Concluding remarks of this section}


Even if we have not exhausted all of the possibilities in each of the above examples, we have shown that one can reach to specific conclusions only after an explicit statement of the postulates that underlay our chosen geometric setup. In other words: there is not ``absolute truth.'' Instead, there are several potential geometric descriptions which are coupled to specific postulates. Only experimentation may rule out those sets of postulates that lead to incorrect phenomenological descriptions.

Another such postulate, that we have not considered here since we have kept our discussion mainly at geometrical level, is related with the gravitational action of given physical theory. This increases the possibilities of describing given gravitational phenomena in a number of different geometrical backgrounds. Actually, consideration of given gravitational action $S_g=\int d^4x\sqrt{-g}{\cal L}_g$, where ${\cal L}_g$ is the gravitational Lagrangian density, adds new possibilities. One may consider, for instance, a gauge invariant gravitational Lagrangian $\sqrt{-g}{\cal L}_g$, so that the derived gravitational equations will respect the manifest symmetry of background space $W_4$. Or one may, alternatively, consider a gravitational Lagrangian without gauge symmetry, even if the geometric background space $W_4$ is gauge symmetric. This, of course, will lead to a gravitational theory that is not gauge invariant, so that the manifest symmetry of the geometric background may be ignored. In this last branch of theories belong the STEGR and STTs of type $f({\cal Q})$.


\section{Do fermions interact with nonmetricity?}\label{sect-fermions}


As we have discussed in the former section \ref{sect-postu}, the role of underlying postulates is essential to get specific results. In the mentioned section we focused in the motion of spinless point particles and showed that the occurrence of the SCE is subject to the assumption of a specific parallel transport law \eqref{ptp} and of a consistency postulate which allows to link physical vectors with related hypothetical vectors in $W_4$ (see subsection \ref{subsect-consist-hypo}). 

Since the geodesics and the autoparallels do not coincide in generalized Weyl space, a question then arises: Does the occurrence of the SCE depend on whether one chooses the geodesics or the autoparallels as the worldlines of test particles? The answer is: Not it does not. The fact that the mass $m$, which is associated with the hypothetical four-momentum ${\bf p}$ \eqref{4-p}, obeys \eqref{master-eq-mass}, is a direct consequence of the nonmetricity law \eqref{n-metricity} and of equation \eqref{cov-der-uvect}, and has nothing to do with whether \eqref{geod-w4} or \eqref{0-geod-w4} determines test particle's worldline. Although there goes a discussion on whether autoparallels or geodesics describe the motion of test particles \cite{adak-arxiv, obukhov}, in this paper we assume that the geodesics are the worldlines of test particles. In this regard, the consistency hypothesis (see subsection \ref{subsect-consist-hypo}) directly relates equation \eqref{master-eq-mass} for mass variation during parallel transport -- more specifically \eqref{mass-law'} -- with timelike geodesic equation \eqref{geod-w4}. 

In any case the geodesics and the autoparallels are irrelevant for the description of the motion of spinor fields (fermions, for instance), as well as of extended spinning test bodies. In the former case it is the Dirac equation in curved spacetime background the one that drives the dynamics, meanwhile, in the latter case the Mathisson-Papapetrou-Dixon equation is the one that describes the motion.


The Lagrangian density of the massless fermion coupled with the gravitational field reads (for simplicity, we omit $SU(2)\otimes U(1)$ gauge terms):

\bea {\cal\hat L}_\text{fermion}=i\bar{\psi}\cancel{\hat D}\psi,\label{psi-lag}\eea where $\psi$ is the Dirac spinor ($\bar{\psi}$ is its adjoint) and the slash gauge derivative is defined as:

\bea \cancel{\hat D}:=\gamma^ae^\mu_a\left(\der_\mu-\frac{1}{2}\sigma_{bc}\hat \omega^{\;\;bc}_\mu+\cdots\right).\label{slash-der}\eea In this equation $\gamma^a$ are the (flat) Dirac gamma matrices, $e^a_\mu$ is the tetrad and the ellipsis stands for the missing terms corresponding to the gauge fields $W^{(i)}_\mu$, $B_\mu$ of the gauge group $SU(2)\otimes U(1)$, which are invariant under Weyl gauge transformations. In \eqref{slash-der} the LC spin connection $\hat\omega^{\;\;ab}_\mu$ and the commutator of the gamma matrices $\sigma^{ab}$ (the generators of the Lorentz group in the spin representation), read:

\bea &&\hat\omega^{\;\;ab}_\mu:=e^{b\nu}\hat\nabla_\mu e^a_\nu=e^{b\nu}\left(\der_\mu e^a_\nu-\{^\lambda_{\mu\nu}\}e^a_\lambda\right),\nonumber\\
&&\sigma^{ab}=\frac{1}{4}\left(\gamma^a\gamma^b-\gamma^b\gamma^a\right),\label{spin-c}\eea respectively. The Lagrangian \eqref{psi-lag} is already gauge invariant, even if in the above form it can be associated with Riemann space $V_4$. This is true also when the $SU(2)\otimes U(1)$ gauge fields $W^{(i)}_\mu$ and $B_\mu$ are included in the Lagrangian density. 

In references \cite{cheng-1988, cheng-arxiv} it has been demonstrated that it does not matter whether one considers the Lagrangian density \eqref{psi-lag} in Riemann space $V_4$ or the equivalent Lagrangian in standard Weyl space $\tilde W_4$: ${\cal L}^*_\text{fermion}$, which amounts to performing the following replacements in \eqref{psi-lag}:

\bea &&\der_\alpha\rightarrow\der^*_\alpha=\der_\alpha+\frac{w}{2}Q^*_\alpha,\;\{^\alpha_{\mu\nu}\}\rightarrow\Gamma^\alpha_{\;\;\mu\nu},\nonumber\\
&&\hat\omega^{\;\;ab}_\alpha\rightarrow\omega^{*\;ab}_\alpha=e^{b\nu}\nabla^*_\mu e^a_\nu\;\Rightarrow\;\cancel{\hat D}\rightarrow\cancel{D}^*,\label{psi-lag-repla}\eea where $w$ is the weight of either $\psi$ or the tetrad $e^a_\alpha$, depending which field the derivative acts on. One gets that 

\bea {\cal\hat L}_\text{fermion}={\cal L}^*_\text{fermion}.\nonumber\eea This means that the vectorial nonmetricity $Q_\alpha$ does not couple neither to fermions nor to other gauge fields including the electromagnetic radiation. 

The demonstration has been extended to arbitrary nonmetricity in \cite{delhom-epjc-2020}, where it has been pointed out that, if take into account an appropriate minimal coupling prescription, the correct Lagrangian should read:

\bea {\cal\hat L}_\text{fermion}=\frac{i}{2}\left[\bar\psi\left(\cancel{\hat D}\psi\right)-\left(\cancel{\hat D}\bar\psi\right)\psi\right].\label{delhom-lag}\eea In other words, fermions and other standard model fields do not interact with nonmetricity, so that the SCE does not take place \cite{cheng-1988, cheng-arxiv, tomi-replay}. 

The above argument is strictly correct only if the mass of the fields $m_\psi$ is assumed vanishing, i. e., if consider the Lagrangian density \eqref{delhom-lag}. In this case the exposed argument is just a confirmation of the result discussed in section \ref{sect-geod}, that photons and radiation interact only with the metric field, i. e., with the LC curvature of spacetime. In other words, that these do not interact with nonmetricity. 

If in place of \eqref{delhom-lag} consider the following Lagrangian density \cite{delhom-epjc-2020}: 

\bea {\cal\hat L}_\text{fermion}=\left\{\frac{i}{2}\left[\bar\psi\left(\cancel{\hat D}\psi\right)-\left(\cancel{\hat D}\bar\psi\right)\psi\right]-\bar\psi m_\psi\psi\right\},\label{psi-mass-lag}\eea where $m_\psi\neq 0$, the conclusion of references \cite{cheng-1988, cheng-arxiv} and related works \cite{tomi-replay}, may be incorrect in general. Actually, if assume the parallel transport law \eqref{ptp} and the consistency hypothesis which, in the present case, amounts to identify the mass parameter $m_\psi$ in \eqref{psi-mass-lag} with the mass in equation \eqref{master-eq-mass} (this is an integral version of equation \eqref{mass-law'}), the mass of the fermion field $m_\psi$ in \eqref{psi-mass-lag} must obey: 

\bea m_\psi(x)=m_\psi(0)\exp\int_{\cal C}\left(Q^*_\lambda+\frac{1}{2}Q_{\lambda\mu\nu}u^\mu u^\nu\right)dx^\lambda,\label{m-ferm-w4}\eea which means that there is a non-negligible (under integral) dependence of the mass $m_\psi$ on nonmetricity in \eqref{psi-mass-lag}, which inevitably leads to the occurrence of the second clock effect. Of course, the identification of the mass of the fermion in \eqref{psi-mass-lag} with the mass in \eqref{m-ferm-w4}, is not an a priori given fact, but it is just a consistency hypothesis, as thoroughly discussed in section \ref{sect-postu} (see also section \ref{sect-geod}, more specifically in subsection \ref{subsect-consist-hypo}).

Notice that the above identification is similar to what we have done with the term $\propto\delta\ln m/\delta x^\alpha$ in the timelike geodesic equation \eqref{geod-w4} which we identified with \eqref{mass-law'} (see section \ref{sect-geod}, more specifically in subsection \ref{subsect-consist-hypo}). Hence, the identification \eqref{m-ferm-w4} is what allows to relate physical vectors and tensors in the gravitational theory, with related (hypothetical) vectors and tensors defined in $W_4$ space.

A similar line of argument may be used in the case of extended spinning test bodies, where in the Mathisson-Papapetrou-Dixon equation, the four-momentum of the (center of mass of the) spinning test body, ${\bf p}$ with coordinate components $p^\mu$, must be identified with the hypothetical four-momentum \eqref{4-p} living in $W_4$. In other words, the mass parameter $m^2=-g_{\mu\nu}p^\mu p^\nu$ must be identified with the one in \eqref{master-eq-mass}. Hence, in agreement with the results for the geodesic motion, timelike spinless point particles, as well as fermion fields and other standard model fields with the mass and extended spinning test bodies, interact with nonmetricity if the above mentioned postulates/hypotheses are assumed to take place. In this case the SCE can not be avoided.


\section{Discussion and conclusion}\label{sect-discuss}


It was known for a long time that Weyl space, that is based on the vectorial nonmetricity condition \eqref{vect-nm}: 

\bea \nabla_\alpha g_{\mu\nu}=-Q_\alpha g_{\mu\nu},\nonumber\eea where $Q_\alpha$ is the Weyl gauge vector, is ruled out as a viable geometrical description of the (classical) laws of gravity, as a consequence of the second clock effect \cite{adler-book, perlick-1991, novello-1992, tucker}. Nevertheless, Weyl geometry has gotten back to scene through proposals to unify the standard model of particles with gravity \cite{smolin-1979, cheng-1988, cheng-arxiv, hochberg-1991, oda-2019, oda-2020, oda-arxiv} and also by looking for applications in cosmology \cite{wei-jcap-2007, maki-2010}, among other interesting physical implications. This is not to mention recent resurrection of Weyl's ideas through a generalization of the nonmetricity condition: $\nabla_\alpha g_{\mu\nu}=-Q_{\alpha\mu\nu}$, which together with the teleparallel condition \eqref{flat-r}, are the geometrical basis of the symmetric teleperallel theories of gravity, including the STEGR. 

STTs and their generalization \cite{nester, adak-2006, adak-2006-1, adak-2013, beltran-plb-2016, javr-prd-2018, vilson-prd-2018, formiga-2019, lavinia-review, beltran-prd-2020, saridakis-prd-2020, lazkoz-prd-2019, lazkoz-prd-2021, sanjay-prd-2020, sanjay-2021, helsinki, 2mann, frusciante, basilakos, lin, mandal, moraes, atayde, dambrosio}, including the so called coincident general relativity \cite{beltran-prd-2018, coincident-1, coincident-2} have been the subject of intensive debate in recent dates. However, little space of the debate has been dedicated to discuss on Weyl gauge symmetry\footnote{In the framework of teleparallel theories of gravity with inclusion of torsion contributions, the issue has been investigated before, for intance, in \cite{maluf-prd-2012, maluf-andp-2012, momeni-2014, silva-2016, silva-2017}.} and the related SCE \cite{2clock-delhom, 2clock-tomi}. As we have shown in section \ref{subsect-ginv-z4}, gauge invariance being a manifest symmetry of $W_4$ space and also of its subclass: teleparallel $Z_4$ space, must play a role in the statement of the STTs. As an example of this, in reference \cite{saridakis-prd-2020}, a class of symmetric teleparallel theories of gravity with gauge symmetry was investigated. In what regards to the related issue about the SCE, in lemma 2 of \cite{2clock-tomi} it was demonstrated that the SCE does not take place in teleparallel spacetimes (symmetric or otherwise). However, the demonstration in \cite{2clock-tomi} is based in equations which are not gauge invariant (see reference \cite{quiros-arxiv-2022} for a discussion of additional inconsistency of the mentioned demonstration).

In this paper we have investigated the consequences of adopting WGS as a manifest symmetry of background spaces with arbitrary nonmetricity. We have shown that if: (i) undertake the gauge invariant parallel transport law \eqref{ptp} and (ii) accept the consistency hypothesis which enables identifying physical vectors and tensors in the given gravitational theory, with related hypothetical vectors and tensors which are defined in $W_4$, the SCE inevitably takes place in generalized Weyl geometry space $W_4$ (also in its subclasses $\tilde W_4$ and $Z_4$). This result invalidates nonmetricity theories -- but for Weyl integrable geometry spaces which are determined by the following choice of nonmetricity: $Q_{\alpha\mu\nu}=\der_\alpha\phi g_{\mu\nu}$ -- as phenomenologically non-viable descriptions of our Universe. In addition, nonmetricity theories predict for the perihelion shift, results that differ from GR predictions. This means that Solar system experiments may rule out nonmetricity theories.


\subsection{Weyl geometry after all}\label{subsect-discuss-b}


Despite of the phenomenological viability argument in connection with the SCE and of the geometric consistency issue, can $W_4$ setup be useful anyway? The answer is: Yes. Let us to discuss about possible scenarios where $W_4$ and the associated gravitational theories, can be viable and even desirable alternatives to Riemann-based gravitational laws.


\subsubsection{Quantum realm}


Perhaps the first application of Weyl's proposal in the quantum domain was presented in \cite{london-1927}. In that work the Weyl gauge field was related with the vector potential of the electromagnetic field of a proton in which an electron was moving: $Q_\alpha=-i(e/\hbar c)A_\alpha$, where $e$ is the electric charge of an electron, $\hbar$ is the reduced Planck constant and $c$ is the speed of light. The length $l$ of any vector associated with the electron, during parallel transport along an electron orbit (a closed curve ${\cal C}$), suffers a variation: $$l=l_0\exp\left(\frac{ie}{2\hbar c}\oint_{\cal C}A_\mu dx^\mu\right).$$ It was required that $$\frac{e}{2\hbar c}\oint_{\cal C}A_\mu dx^\mu=2\pi n,$$ so that in an orbit $l=l_0$. This led to the Bohr quantization rule in connection with the allowed radii in hydrogen \cite{adler-1970}. It was subsequently demonstrated that the Weyl gauge vector can not be the vector potential of the electromagnetic field, so that this work was forgotten.\footnote{It is interesting to notice that, thanks to the Stokes theorem: $$\oint_{\cal C}A_\mu dx^\mu=\frac{1}{2}\int_\Sigma F_{\mu\nu}d\sigma^{\mu\nu},$$ where $F_{\mu\nu}=\der_\nu A_\mu-\der_\mu A_\nu$ is the electromagnetic tensor, $\Sigma$ is an oriented surface bounded by the curve ${\cal C}$, and $d\sigma^{\mu\nu}$ is the infinitesimal element of surface. Means that the flux of the electromagnetic field $\Phi_F=\int_\Sigma F_{\mu\nu}d\sigma^{\mu\nu}/2$ obeys the ``quantization condition:'' $\Phi_F=4\pi n\hbar c/e$.} 

In reference \cite{wheeler-1990} it is demonstrated that a complete theory of measurement in a Weyl geometry contains the crucial elements of quantization, so that independent introduction of operators and commutation relations into this geometry is unnecessary. In this regard it is shown that quantization and uncertainty of measurement arise in a natural way from certain assumptions about the nature of motion in a Weyl geometry. Other attempts at quantization in the framework of Weyl geometry are based in the de Brooglie-Bohm approach to quantum mechanics \cite{wood-1997, arabs-2003}. While in the model presented in \cite{wood-1997} Weyl geometry is used to express the principles of the causal interpretation of quantum mechanics in a fully relativistic form, in \cite{arabs-2003} a Weyl invariant theory is built, and it is shown that both gravity and quantum are present at the level of equations of motion. The authors conclude that the natural framework for both gravity and quantum is Weyl geometry.\footnote{There has been recent attempts at the geometrization of quantum mechanics \cite{castro-2021}.} In \cite{godfrey-1984} a formalism was proposed that was based on the assumption that the underlying geometric structure of the phase space (instead of spacetime) is that of Weyl geometry. Under this assumption the Dirac commutator emerges. From this formalism quantum mechanics can be deduced.  

In general terms, Weyl geometry framework is by far a most adequate formalism to address the quantization of gravity than Riemann geometry. This is related with built-in gauge invariance in Weyl spaces and the associated variation of the length of vectors during parallel transport. If take advantage of these features of Weyl geometry spaces, there are much more possibilities to approach quantization than in Riemann spaces. When gravity is incorporated into the scheme of quantum field theory, the spacetime itself is dynamical in the sense that it is affected by, and also affects, the fields it contains. The geometry itself is subject to quantum fluctuations. In this regard Weyl geometry, where the measurement process is affected by the gravitational interactions, seems to offer a very attractive scenario for quantization.


\subsubsection{Gauge invariant gravitation}


An aspect of theories with gauge invariance that is usually avoided is related with the underlying geometric structure of spacetime. For instance in \cite{bars-turok}, where a gauge invariant theory of gravity and of the standard model of particles (SMP) is developed, nothing is said about this issue. However, it is implicit that the geometric background is Riemann space $V_4$ in that reference. In such a case, the action and the geometric setup do not share the same symmetry (gauge symmetry in the present case). An alternative point of view on gauge invariance is developed in references \cite{dirac-1973, smolin-1979, cheng-1988}, where the gauge invariant action is associated with Weyl geometry background spaces $\tilde W_4$. In all of these works break down of gauge symmetry is required in order to meet phenomenological viability. 

A different point of view is put forth in \cite{quiros-arxiv}, where a gauge invariant theory of gravity and of the SMP based in Weyl integrable space, was investigated. The theory is based based on the following postulates:

\begin{enumerate}

\item The affine structure of the spacetime is determined by Weyl-integrable geometry and not by the laws of Riemann geometry.

\item Only the fields (this includes the metric, the gauge scalar, the mass, the Higgs, the spinors and other matter fields) are transformed by the Weyl rescalings. The dimensionless as well as the dimensionful constant parameters, including the fundamental constants of nature such as the Planck constant $\hbar$, the speed of light, the fine structure constant, $\alpha$, etc., are not transformed by the Weyl gauge transformations.

\end{enumerate} Actual dimensionless and dimensionful constants -- called as ``bare constants'' -- are not transformed by the gauge transformations, in contrast to point dependent constants which are obtained as the product of a bare constant by an appropriate power of the (exponent of the) gauge scalar. In order to incorporate the standard model of elementary particles in a gauge invariant way into the theory, it suffices to promote the electroweak (EW) mass parameter $v_0$ to a point dependent quantity \cite{bars-turok}: $v_0\rightarrow v=v_0\exp{(\vphi/2)}$. We want to stress that the only way in which the standard unit of mass can be a point dependent quantity, as required by Weyl geometric laws, is that the particles of the SMP acquired point dependent masses as a result of the EW symmetry breaking procedure. This is what is achieved by the theory \cite{quiros-arxiv}. In this theory, as a result of EW symmetry breaking the Higgs acquires a point dependent VEV; $|H|=v_0\,e^{\vphi/2},$ so that the gauge bosons and fermions of the SMP acquire point-dependent masses. In particular Fermion fields $\psi$, acquire point dependent masses through Yukawa interactions of the form: $g_\psi\bar\psi H\psi$ ($g_\psi$ is a Yukawa coupling). Hence, the mass of the fermion: $m_\psi=g_\psi v_0\,e^{\vphi/2},$ is a point dependent quantity as well. The resulting gauge invariant theory of gravity and of the SMP, preserves the gauge symmetry even after $SU(2)\otimes U(1)$ symmetry breaking. This means that if this were the correct (classical) theory of gravity, gauge symmetry could be an actual symmetry of the laws of physics in our present Universe.


\subsubsection{From minimum length to cosmology: the span of gauge symmetry}


According to one of the most popular approaches to quantum gravity, there exists a minimal length scale (or a maximum energy scale) that plays a fundamental role in the laws of nature \cite{garay, sabine}. In reference \cite{sabine}, a quite recent review of models that have been developed to implement a minimal length scale in quantum mechanics and quantum field theory, is performed. These models have entered the literature as the generalized uncertainty principle or the modified dispersion relation, and have fed the study of the effects of a minimal length scale in quantum mechanics, quantum electrodynamics, thermodynamics, etc.

Let us wonder, what does the existence of a minimum length entail for gauge symmetry? The first thing that jumps out is that, following the most widespread reasoning line, in a quantum world with minimum length Weyl gauge invariance, if exists, should be a broken symmetry due to the existence of an intrinsic scale associated with this minimum length (let us assume it is Planck length). But, if follow the same reasoning line, after EW symmetry breaking, due to existence of EW mass scale, gauge symmetry should be also a broken symmetry. This means that gauge symmetry could be a symmetry of the physical laws in the period between the Planck and EW scales. This is a period where the Universe is filled with radiation, but radiation does not interact with the only geometrically consistent Weyl nonmetricity \eqref{vect-nm}. Hence, this widespread line of reasoning leaves little room for gauge symmetry to be a fundamental symmetry in Nature. Alternatively, if one insists in describing cosmology within a gauge invariant geometrical framework, as it is done in recent works on symmetric teleparallel cosmology \cite{capozziello-2016, coley-2019, beltran-plb-2016, beltran-prd-2020, saridakis-prd-2020, lazkoz-prd-2019, lazkoz-prd-2021, sanjay-prd-2020, sanjay-2021, frusciante, basilakos, mandal, atayde}, gauge symmetry must survive EW symmetry breaking so that it can be a symmetry of our present (and perhaps future) Universe.


\section*{Acknowledgments}

The author thanks M P Hobson, A N Lasenby and M Adak for their useful comments and T Koivisto and L Heisenberg for pointing to us lemma 2 of reference \cite{2clock-tomi} and for mentioning several bibliographic references. Support of this research by FORDECYT-PRONACES-CONACYT under grant CF-MG-2558591, is deeply acknowledged.


\appendix



\section{Equations of motion of particles with point-dependent mass}\label{app-a}


Let us write the action for a point-particle of mass $m$ moving in $W_4$ \eqref{action-geod-w4}:

\bea S=\int mds,\label{app-action}\eea where, since under \eqref{gauge-t}: $ds\rightarrow\Omega ds$ and $m\rightarrow\Omega^{-1}m$, the action is gauge invariant. If one applies to this action the standard variational principle of the least action, we get:

\bea &&\delta S=\int\left[\delta mds+m\delta(ds)\right]=\nonumber\\
&&\int\delta x^\alpha mds\left[\frac{\delta\ln m}{\delta x^\alpha}-g_{\alpha\lambda}\frac{d^2x^\lambda}{ds^2}-g_{\alpha\lambda}\frac{\delta\ln m}{\delta x^\kappa}\frac{dx^\kappa}{ds}\frac{dx^\lambda}{ds}\right.\nonumber\\
&&\left.\;\;\;\;\;\;\;\;\;\;\;-\frac{1}{2}\left(\der_\kappa g_{\lambda\alpha}+\der_\lambda g_{\kappa\alpha}-\der_\alpha g_{\kappa\lambda}\right)\frac{dx^\kappa}{ds}\frac{dx^\lambda}{ds}\right],\label{app-var}\eea where we took into account that $$m\delta(ds)=m\delta\sqrt{g_{\kappa\lambda}dx^\kappa dx^\lambda},$$ and we omitted a total derivative $d[g_{\kappa\lambda}(dx^\kappa/ds)\delta x^\lambda]$ under the integral. In the process of derivation we have assumed that:

\bea \delta g_{\mu\nu}=\der_\lambda g_{\mu\nu}\delta x^\lambda,\;dg_{\mu\nu}=\der_\lambda g_{\mu\nu}dx^\lambda.\nonumber\eea The equation of motion \eqref{geod-w4} that is obtained from \eqref{app-var} by requiring that $\delta S=0$, reads:

\bea \frac{d^2x^\alpha}{ds^2}+\{^\alpha_{\mu\nu}\}\frac{dx^\mu}{ds}\frac{dx^\nu}{ds}+\frac{\delta\ln m}{\delta x^\mu}h^{\alpha\mu}=0,\label{app-geod-1}\eea where the orthogonal projection tensor $h^{\alpha\beta}$ is defined in \eqref{proj-t} and we left $\delta\ln m/\delta x^\mu$, to mean that, in general, it is not a partial derivative since $d\ln m$, is not a perfect differential. The obtained geodesic equation is manifestly gauge invariant.




\end{document}